\documentclass[aip,jcp,reprint,graphicx,twocolumn,amsmath,amssymb,groupedaddress]{revtex4-1}

\usepackage{bm}
\usepackage[normalem]{ulem}
\usepackage{epsfig}
\usepackage{graphicx}
\usepackage{amssymb,amsmath,amsbsy,amsgen,amsfonts}    
\usepackage{dcolumn}
\usepackage{amsthm}
\usepackage{mathrsfs}
\usepackage{latexsym}
\usepackage{array}
\usepackage{amstext}
\allowdisplaybreaks[1]
\usepackage{txfonts}

\newcommand{\bra}[1]{\ensuremath{\left\langle{#1}\right\vert}}

\newcommand{\ket}[1]{\ensuremath{\left|{#1}\right\rangle}}

\newcommand{\abs}[1]{\left|#1\right|}
\def\bea{\begin{eqnarray}}
\def\eea{\end{eqnarray}}

\def\bea{\begin{eqnarray}}
\def\eea{\end{eqnarray}}

\makeatother

\begin{document}

\title{Vibronic origin of long-lived coherence in an artificial molecular light harvester}

\author{James Lim$^{1,\ast}$, David Pale\v{c}ek$^{2,3,\ast}$, Felipe Caycedo-Soler$^{1}$, Craig N. Lincoln$^{4}$, Javier Prior$^{5}$, Hans von Berlepsch$^{6}$, Susana F. Huelga$^{1}$, Martin B. Plenio$^{1}$, Donatas Zigmantas$^{2}$, and J\"{u}rgen Hauer$^{4}$}
\affiliation{
$^1$ Institut f{\"u}r Theoretische Physik, Albert-Einstein Allee 11, Universit{\"a}t Ulm, 89069 Ulm, Germany\\
$^2$ Department of Chemical Physics, Lund University, P.O. Box 124, SE-22100 Lund, Sweden\\
$^3$ Department of Chemical Physics, Charles University in Prague, Ke Karlovu 3, 121 16 Praha 2, Czech Republic\\
$^4$ Photonics Institute, Vienna University of Technology, Gusshausstrasse 27, 1040 Vienna, Austria\\
$^5$ Departamento de F{\'i}sica Aplicada, Universidad Polit{\'e}cnica de Cartagena, Cartagena 30202, Spain\\
$^6$ Forschungszentrum f{\"u}r Elektronenmikroskopie, Institut f{\"u}r Chemie und Biochemie, Freie Universit{\"a}t Berlin, Fabeckstra$\beta$e 36a, D-14195 Berlin, Germany\\
$^\ast$ These authors contributed equally to this work.}


\begin{abstract}
Natural and artificial light harvesting processes have recently gained new interest. Signatures of long lasting coherence in spectroscopic signals of biological systems have been repeatedly observed, albeit their origin is a matter of ongoing debate, as it is unclear how the loss of coherence due to interaction with the noisy environments in such systems is averted. Here we report experimental and theoretical verification of coherent exciton-vibrational (vibronic) coupling as the origin of long-lasting coherence in an artificial light harvester, a molecular J-aggregate. In this macroscopically aligned tubular system, polarization controlled 2D spectroscopy delivers an uncongested and specific optical response as an ideal foundation for an in-depth theoretical description. We derive analytical expressions that show under which general conditions vibronic coupling leads to prolonged excited-state coherence.
\end{abstract}

\maketitle


{\bf Introduction}

The remarkably high efficiency in photosynthesis,
where nine out of ten absorbed photons reach the reaction center,
is a fascinating field of modern research. In such photosynthetic
complexes, structure, dynamics and function are inextricably linked.
A conserved building block comprises strongly absorbing pigments arranged
in close proximity to one another by a surrounding protein scaffold\cite{van_Grondelle,Blankenship}.
Typical inter-pigment distances are of order of $10\,\AA$ and photon
absorption leads to the formation of delocalized excited electronic
states (excitons) shared by two or more pigment molecules. Exciton
creation, migration and trapping are central to the functionality
of a photosynthetic apparatus. The controlled and adjustable arrangement
of the pigments tunes the electronic network and the properties of
its interaction with the vibrational environment that is associated
with either the pigments or the protein. The detailed balance of these
properties determines the efficiency of light harvesting systems\cite{Renger2001,HuelgaPlenio_CP2013}.

Exciton dynamics can be efficiently probed by two-dimensional (2D)
electronic spectroscopy\cite{Jonas_ARPC2003}. This technique revealed oscillatory signals in the spectral response of a wide variety
of photosynthetic aggregates\cite{Engel_Nature2007,Dostal2014}. Initially
ascribed to excitonic beatings, oscillations have been found to persist
up to several hundreds of femtoseconds at room temperature\cite{Collini_Nature2010,Romero_NaturePhys2014,Fuller_NatureChem2014}.
This time scale exceeds typical dephasing rates in the condensed phase
and becomes comparable to exciton transfer times\cite{van_Grondelle},
thus posing the question of the nature and functional relevance
of these coherences \cite{HuelgaPlenio_CP2013}. Unfortunately, the
complex structure of 2D signals makes the unambiguous identification
of the underlying mechanisms that support such long-lived coherences
a challenging task and several hypotheses to explain them have been formulated \cite{Chin_NaturePhys2013,Plenio_JCP2013,ChinHP12,Kolli_JCP2012,Jonas_PNAS2012,Fleming_Science2007,Ishizaki_PCCP2010,Engel_Science2013,Christensson_JPCB2011,Caycedo_JCP2012,Christensson_JPCB2012}.
The different approaches can be classified into theories including
coherent interaction of excitons with intra-pigment vibrations\cite{Chin_NaturePhys2013,Plenio_JCP2013,ChinHP12,Kolli_JCP2012,Jonas_PNAS2012}
and theories focusing on incoherent exciton-protein interaction such
as correlated fluctuations\cite{Fleming_Science2007,Ishizaki_PCCP2010,Engel_Science2013}.
It is possible that some of these mechanisms may coexist on certain
time scales and that one or another may become dominant depending
on the system under consideration.

In this work, we show that the relatively simple excitonic structure of a molecular J-aggregate provides an ideal test case to identify the microscopic mechanism behind long-lived oscillations in electronic 2D-signals. The investigated J-aggregate is tubular and aligns along the sample's flow direction when in solution. Additionally, the J-aggregate exhibits excitonic bands with roughly orthogonal transition dipole moments. It is this combination of perpendicular excitonic transitions and macroscopic alignment that makes electronic 2D-spectroscopy with polarization-controlled excitation pulses an ideal tool to study coherence effects between the excitonic bands. This approach significantly reduces the complexity of retrieved 2D signals, leading to only two peaks with oscillatory components in specific regions of the 2D-maps, {\it i.e.} one on the diagonal and one as a cross-peak for non-rephasing and rephasing signal components, respectively. Employing a vibronic model, we derive analytical expressions that show how system parameters such as electronic decoherence rates and exciton-vibrational resonance determine the amplitude and lifetime of oscillatory signals. Fitting the analytical expressions to measured data, the vibronic model achieves quantitative agreement with experimental observations. Concerning potential functional relevance of the observed oscillations, we show that the long-lived oscillatory signals in our system are dominated by excited-state coherence rather than ground-state coherence.


\

{\bf Results}


{\bf The system.} J-aggregates of cyanine dyes are promising candidates
for artificial antenna systems\cite{Heijs_CP2007,Wurthner_ACIE2011,Eisele_PNAS2014,Yuen-Zhou_ACSNano2014,Qiao_ACSNano2015}.
They are chemically versatile and self-assemble into various extended
supramolecular structures in aqueous solution\cite{Berlepsch}. Here
a system that can be considered a macroscopically aligned synthetic
light harvester was studied, namely a molecular J-aggregate of C8O3-monomers
whose aggregation behavior is well known\cite{Berlepsch2002,Berlepsch2003}.
As revealed by cryogenic transmission electron microscopy\cite{Berlepsch_JPCB2000},
the aggregate structure is best described as a double-layered nanotube
with outer diameter $\sim$$\,11\,{\rm nm}$ and lamellar spacing of $\sim$$\,2.2\,{\rm nm}$
between the chromophore layers. Additionally, superhelical bundles
of these tubes can also form, though the addition of polyvinyl alcohol
(PVA) inhibits this process and thereby avoids single-layered tube formation\cite{Eisele_PNAS2014}
and maintains a stable solution over several weeks\cite{Berlepsch_JPCB2003}.
A drawing of the J-aggregate under investigation, from here on referred
to as C8O3, is shown in Fig.1\textbf{a}. The bilayer configuration
of C8O3 allows the effect of different decoherence rates to be studied
as the outer solvent-exposed layer shows faster decoherence than the
inner protected layer.

\begin{figure}
\includegraphics{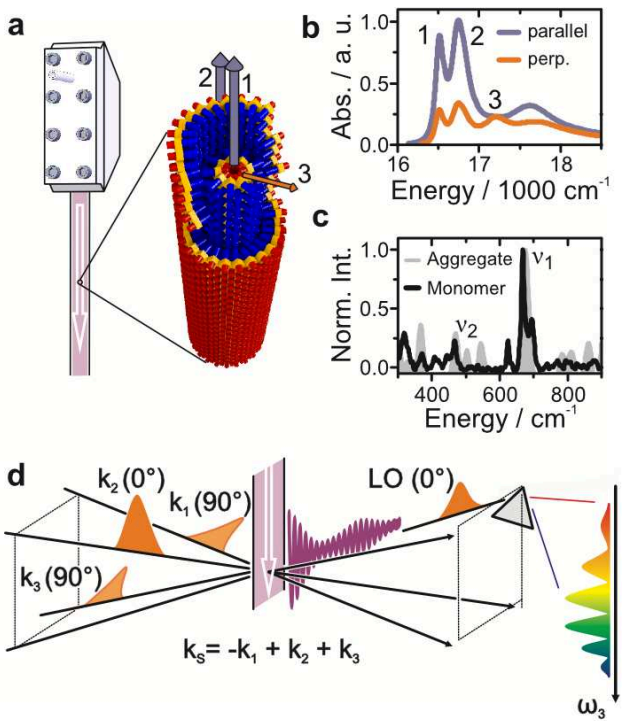} \protect\caption{\textbf{C8O3 and polarization controlled 2D spectroscopy}. \textbf{a},
Wire-guided window-free jet used for sample circulation, along with
a schematic of the double-layered structure of the C8O3-aggregate. The aggregates align along the flow direction (white arrow). The transition dipole directions of bands 1-3 are displayed by arrows, which are mainly polarized along the tube axis (bands 1 and 2 shown in blue) or perpendicular to the axis (band 3 shown in orange). \textbf{b}, Absorption
spectra with light polarized parallel (blue) and perpendicular (orange)
to the flow direction. \textbf{c}, Non-resonant Raman spectra of the
C8O3-monomer (black line) and aggregate (grey area). The vibrational
frequencies $\nu_{1}$ and $\nu_{2}$ are close to the exciton energy
splitting between bands 1 and 3 and bands 2 and 3, respectively. \textbf{d},
Polarization controlled 2D spectroscopy with three excitation pulses
($k_{1}$ to $k_{3}$) and a local oscillator (LO) for heterodyne
detection of the signal field, depicted as an oscillating line. Polarization
orientation ($0^{\circ}$ or $90^{\circ}$) is given with respect
to the longitudinal axis of aligned C8O3.}
\label{figure1} 
\end{figure}

The structural properties of the aggregate are remarkable: the $11\,{\rm nm}$
outer diameter is contrasted by a length of several micrometers. Circulating
solvated C8O3 with a wire-guided jet (Fig.1\textbf{a}) leads to a
macroscopic orientation of the tubes because the longitudinal axis
preferentially aligns along the flow direction. This creates anisotropy
for linearly polarized light, as shown in Fig.1\textbf{b}. Linear
dichroism measurements\cite{Berlepsch_JPCB2003} and redox-chemistry
studies\cite{Eisele2012} assign bands 1 and 2 to longitudinal transitions
localized upon the inner and outer cylinders, respectively (Fig.1\textbf{a}).
Transitions to band 3 are preferentially polarized perpendicular to
the long axis of C8O3 and are shared by both layers. A detailed description
of sample preparation methods and band assignments is given in the
Supplementary Notes 1 and 2.

Fitting the well-defined absorption peaks of C8O3 with Lorentzian
functions (see Supplementary Note~2) reveals an exciton energy difference between bands
1 and 3 of $\Delta\Omega_{31}\approx690\,{\rm cm}^{-1}$ and $\Delta\Omega_{32}\approx460\,{\rm cm}^{-1}$
for bands 2 and 3. Both exciton energy splittings are close to vibrational
frequencies $\nu_{1}\approx668\,{\rm cm}^{-1}$ and $\nu_{2}\approx470\,{\rm cm}^{-1}$
observed in non-resonant Raman spectra\cite{Milota_JPCA2013} (Fig.1\textbf{c}).
These vibrational frequencies are measured in both the monomer and
aggregate Raman spectra, {\it i.e.}~they are not aggregation induced Raman
bands. Strongly enhanced modes at similar energies were observed in
resonant Raman spectra of a related cyanine dye, and can be assigned
to out-of-plane vibrations\cite{Aydin_JCP2011}. Such out-of-plane
vibrations were shown to couple strongly to excitons\cite{Rich2013}.
The quasi-resonance between the vibrational frequencies $\nu_{1}$
and $\nu_{2}$ and exciton energy splittings $\Delta\Omega_{31}$
and $\Delta\Omega_{32}$ provides us with an interesting scenario
of possible coherent interaction between bands (excitons) and vibrations\cite{Chin_NaturePhys2013,ChinHP12,Kolli_JCP2012,Christensson_JPCB2012,Butkus2013}.
Such exciton-vibrational coupling induces vibronic\cite{Plenio_JCP2013}
and vibrational coherences\cite{Jonas_PNAS2012}, which can both
lead to long-lived beating signals in 2D spectra. Here we emphasize
that coherence in the electronic excited-state manifold is referred to as vibronic
and in the ground-state manifold as vibrational. Identifying the dominant contribution
is of fundamental importance because only vibronic coherence, which
manifests in excited state dynamics, can enhance exciton transport
and thus support light-harvesting function\cite{Womick_JPCB2009,Womick_JPCB2011,delReyCH+13}.


{\bf Experimental results.} The absorption spectrum of a light
harvesting system may be heavily congested because of overlapping
excitonic bands and the resulting 2D-signal would exhibit significant
overlap between diagonal and cross peaks, thereby impeding further
analysis. It has been suggested to employ laser pulses of different
relative polarization to selectively address relevant excitation pathways
to obtain a clearer 2D signal\cite{Hochstrasser_CP2001}. However,
the advantage of polarization controlled 2D spectroscopy has been
limited by the isotropic nature of the investigated samples (an ensemble).
In the experiment presented here, these problems are circumvented
by the measurement of the macroscopically aligned C8O3. The transition
dipole moments of bands 1 and 2 are preferentially parallel to the
longitudinal axis while band 3 is orthogonal, thus allowing for optimal
polarization selectivity. This combination reduces the obtained 2D
maps to only two relevant peaks with negligible overlap and an up
to 30 times stronger signal intensity as compared to the isotropic
case\cite{Read_PNAS2007}.

\begin{figure}[!tbph]
\includegraphics{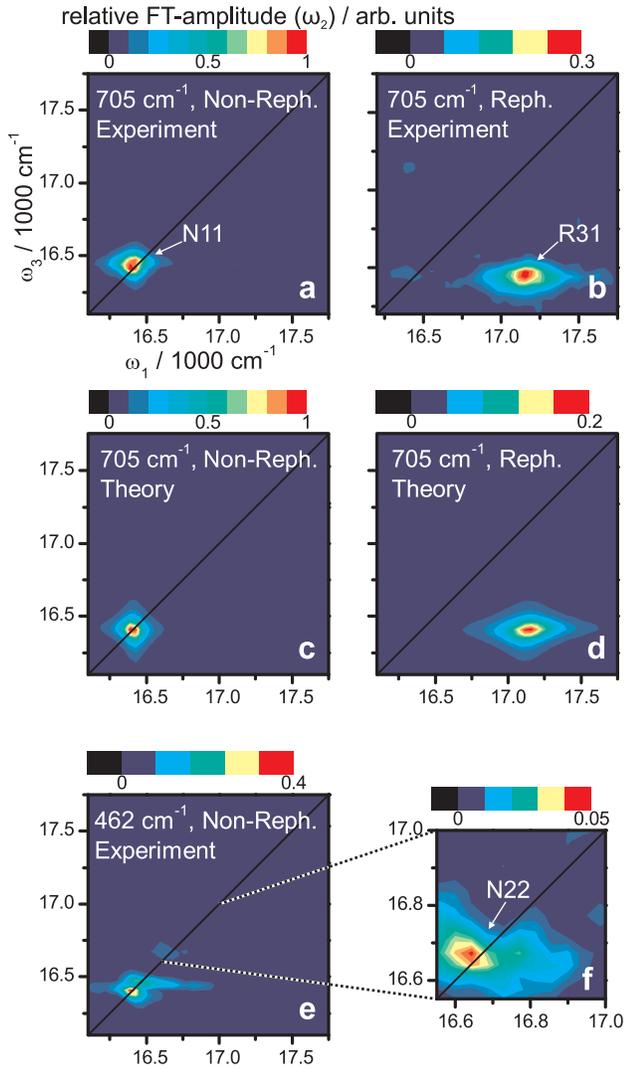} \protect\caption{\textbf{Experimental and theoretical 2D spectra}. \textbf{a},\textbf{
b}, The Fourier-transform amplitude maps of non-rephasing and rephasing
spectra at $\omega_{2}=705\pm20\,{\rm cm}^{-1}$, which reveal the
presence of a non-rephasing diagonal peak N11 and a rephasing cross-peak
R31. These peaks stem from the coherent interaction of bands 1 and
3 with the quasi-resonant vibrational mode with frequency $\nu_{1}\approx668\,{\rm cm}^{-1}$.
The amplitude of N11 is about three times larger than R31. The lineshape
of N11 is symmetric along both $\omega_{1}$- and $\omega_{3}$-axes,
while that of R31 is elongated along $\omega_{1}$-axis. \textbf{c},\textbf{
d}, The simulated spectra at $\omega_{2}=705\,{\rm cm}^{-1}$ with
N11 and R31. \textbf{e}, The FT amplitude map at $\omega_{2}=462\pm20\,{\rm cm}^{-1}$
reveals coherent interaction of bands 2 and 3 with the quasi-resonant
vibrational mode with frequency $\nu_{2}\approx470\,{\rm cm}^{-1}$.
However, as depicted in \textbf{f}, the associated non-rephasing peak
N22 at $\omega_{1,3}\approx16670\thinspace{\rm cm^{-1}}$ is weak
and only amounts to 5\% of N11 at $\omega_{2}=705\pm20\,{\rm cm}^{-1}$
(see \textbf{a}). The diagonal peak at $\omega_{1,3}\approx16400\thinspace{\rm cm^{-1}}$
in \textbf{e} stems from N11, with a peak centered at $\omega_{2}=705\pm20\,{\rm cm}^{-1}$,
but broad enough to appear at $\omega_{2}=462\pm20\thinspace{\rm cm}^{-1}$.
All measurements were carried out at room temperature.}

\label{figure2} 
\end{figure}

The ideal pulse sequence to isolate beating signals between states
with orthogonal transition dipole moments, \textit{i.e.}~bands 1 and
3 in the present case, is depicted in Fig.1\textbf{d}, where the phase-matched direction for measuring rephasing spectra is displayed: non-rephasing spectra can be measured along the same phase-matched signal direction by changing the order of the first two pulses (see Methods).
After subtraction of the non-oscillatory background, we performed
a Fourier transformation along waiting time $t_{2}$ for all points
on the two-dimensional $(\omega_{1},\omega_{3})$-map. The resulting
$\omega_{2}$-plots allow the lineshape of beating signal with frequency
$\omega_{2}$ to be visualized as a function of position in $(\omega_{1},\omega_{3})$-space.
The slice at the exciton energy splitting between bands 1 and 3 ($\omega_{2}=705\pm20\,{\rm cm}^{-1}$
with the experimental resolution of $\pm20\,{\rm cm}^{-1}$) reveals
a non-rephasing diagonal peak N11 and a rephasing cross-peak R31 as
shown in Figs.2\textbf{a} and \textbf{b}, respectively. N11 is centered
at $(\omega_{1},\omega_{3})=(\Omega_{1},\Omega_{1})$ with exciton
energy $\Omega_{1}\approx16405\,{\rm cm}^{-1}$ of band 1 and a symmetric
linewidth $2\Gamma_{g1}\approx130\,{\rm cm}^{-1}$ along both $\omega_{1}$-
and $\omega_{3}$-axes (Fig.2\textbf{a}). The center of R31 is located
at $(\omega_{1},\omega_{3})=(\Omega_{3},\Omega_{1})$ with exciton
energy $\Omega_{3}\approx17125\,{\rm cm}^{-1}$ of band 3 and asymmetric
linewidths $2\Gamma_{g3}\approx300\,{\rm cm}^{-1}$ and $2\Gamma_{g1}\approx130\,{\rm cm}^{-1}$
along $\omega_{1}$- and $\omega_{3}$-axes, respectively (Fig.2\textbf{b}).
In peak amplitude, R31 is approximately $30\,\%$ of N11. Turning
to the $\omega_{2}$-slice corresponding to the energy splitting between
bands 2 and 3, ($\omega_{2}=462\pm20\,{\rm cm}^{-1}$), Figs.2\textbf{e}
and \textbf{f} reveal a diagonal non-rephasing peak N22, which is
centered at $(\omega_{1},\omega_{3})=(\Omega_{2},\Omega_{2})$ with
the exciton energy $\Omega_{2}\approx16672\,{\rm cm}^{-1}$ of band
2 and a symmetric linewidth $2\Gamma_{g2}\approx225\,{\rm cm}^{-1}$
along $\omega_{1}$- and $\omega_{3}$-axes. The amplitude of N22
is only $5\,\%$ of N11.


{\bf Theoretical model.} In order to describe the long-lived oscillations
in N11 and R31, a vibronic model is employed that describes the coupling
of bands 1 and 3 to a quasi-resonant vibrational mode with frequency
$\nu_{1}$. Consider a system with electronic ground state $\ket{g_{k}}$
and excited states for bands 1 and 3, denoted by $\ket{1_{k}}$ and $\ket{3_{k}}$, respectively, 
where $k=0$ and 1 denote the \emph{vibrational} ground and excited
state, respectively (Fig.3\textbf{a}). The vibronic coupling between
the quasi-resonant states $\ket{3_{0}}$ and $\ket{1_{1}}$ leads
to unnormalized vibronic eigenstates $\mathinner{\langle\tilde{3}_{0}|}=\bra{3_{0}}+\epsilon\bra{1_{1}}$
and $\mathinner{\langle\tilde{1}_{1}|}=\bra{1_{1}}-\epsilon\bra{3_{0}}$.
Here, $\epsilon$ represents the degree of vibronic mixing defined
by 
\begin{equation}
\epsilon=i\nu_{1}\sqrt{S_{1}}(i\Delta\nu_{1}-\Gamma_{13})^{-1},\nonumber
\end{equation}
where $\Delta\nu_{1}=(\Omega_{3}-\Omega_{1})-\nu_{1}$ denotes the
detuning between $\ket{3_{0}}$ and $\ket{1_{1}}$, \textit{i.e.}~between
the exciton energy splitting and vibrational frequency, and $S_{1}$
denotes the Huang-Rhys factor of the vibrational mode, which in turn
quantifies the strength of the vibronic coupling (see Supplementary Note~2 for details
of the derivation). The electronic decoherence rate $\Gamma_{gk}$
describes the exponential decay rate of the coherence between electronic
ground state and band $k$, while $\Gamma_{13}$ represents the overall
exponential decay rate of the inter-exciton coherence between bands
1 and 3. In our model, we do not consider inhomogeneous broadening,
which is justified by the observation that the experimentally measured
absorption spectrum is well matched to a sum of Lorentzian functions
with the linewidths $2\Gamma_{gk}$ (see Supplementary Note~2). This is valid when homogeneous
broadening dominates the linewidths and the Huang-Rhys factors are
sufficiently small, as is the case here. In addition, the lineshape
of N11 (Fig.2\textbf{a)} is not elongated along the diagonal $\omega_{1}=\omega_{3}$,
implying our 2D signal is dominated by homogeneous broadening. The
same conclusion is reached from analyzing 2D correlation spectra\cite{Milota_JPCA2013}.

In nonlinear spectroscopy, the molecular response to laser excitation
is described by response functions\cite{Mukamel}. According to the
vibronic model described above, the response function for the oscillatory
signals in N11 reads 
\begin{equation}
{\cal R}_{N11}=\mu_{1}^{2}\mu_{3}^{2}\Gamma_{g1}^{-2}(e^{[i(\Delta\Omega_{31}+\delta\omega)-\Gamma_{13}]t_{2}}+e^{[i(\nu_{1}-\delta\omega)-\gamma_{v}]t_{2}}\epsilon^{2}),\nonumber
\end{equation}
with $\mu_{1}$ and $\mu_{3}$ denoting the transition dipole moment
of bands 1 and 3, respectively. The prefactor $\Gamma_{g1}^{-2}$
stems from the lineshape of N11, $\gamma_{v}$ denotes the dissipation
rate of the vibrations and $\delta\omega$ stands for the frequency
shift of the vibronic eigenstates $\mathinner{\langle\tilde{3}_{0}|}$
and $\mathinner{\langle\tilde{1}_{1}|}$ relative to the uncoupled
states $\bra{3_{0}}$ and $\bra{1_{1}}$ due to the vibronic coupling
(see Fig.3\textbf{a} and Supplementary Note~2 for further details). The coupling was
found to be sufficiently strong to induce non-negligible vibronic
mixing $|\epsilon|^{2}\approx0.03$, which leads to a long-lived beating
signal in N11 up to $t_{2}\approx800\,{\rm fs}$, as shown in Fig.3\textbf{b}.
These results imply that the initial excitonic part of $\ket{1_{0}}\mathinner{\langle\tilde{3}_{0}|}$
decays rapidly with 1/e decay time of $\Gamma_{13}^{-1}\approx66\,{\rm fs}$,
while the vibronic coherence $\ket{1_{0}}\mathinner{\langle\tilde{1}_{1}|}$
explains a long-lived oscillatory signal in N11: here $\ket{1_{0}}\mathinner{\langle\tilde{3}_{0}|}$
($\ket{1_{0}}\mathinner{\langle\tilde{1}_{1}|})$ represent coherence
between two vibronic states $\ket{1_{0}}$ and $\mathinner{\langle\tilde{3}_{0}|}$
($\ket{1_{0}}$ and $\mathinner{\langle\tilde{1}_{1}|}$), respectively.

The response function for the oscillatory contributions to R31 is
given by 
\begin{equation}
{\cal R}_{R31}=\mu_{1}^{2}\mu_{3}^{2}\Gamma_{g3}^{-1}\Gamma_{g1}^{-1}(e^{[i(\Delta\Omega_{31}+\delta\omega)-\Gamma_{13}]t_{2}}+e^{[i(\nu_{1}-\delta\omega)-\gamma_{v}]t_{2}}\epsilon^{2}(\eta_{e}-\eta_{g})),\nonumber
\end{equation}
where $\Gamma_{g3}^{-1}\Gamma_{g1}^{-1}$ derives from the asymmetric
lineshape of R31 (see Figs.2\textbf{b} and \textbf{d}). Here $\eta_{e}$
and $\eta_{g}$ represent the contribution of excited-state vibronic
coherence $\ket{1_{0}}\mathinner{\langle\tilde{1}_{1}|}$ and ground-state
vibrational coherence $\ket{g_{0}}\bra{g_{1}}$, respectively, to
the long-lived beating signal in R31 (see Supplementary Note~2). The vibrational coherence
in the electronic ground-state manifold does not play a role in exciton
transfer dynamics, but nonetheless modulates the 2D spectra. A fit
of model parameters to experimental results (Fig.3\textbf{c}) shows
that $|\eta_{e}|\approx2.5\,|\eta_{g}|$. This means the long-lived
beating signal in R31 is dominated by the excited-state coherence
$\ket{1_{0}}\mathinner{\langle\tilde{1}_{1}|}$. The short-lived beating
signal in R31 is induced by $\ket{1_{0}}\mathinner{\langle\tilde{3}_{0}|}$,
as is the case for N11. We note that the signal at N11, with approximately
three times the amplitude of R31, is exclusively determined by excited-state
contributions. Details of this vibronic model and the corresponding
Feynman diagrams for the spectral components N11 and R31 are discussed
in the Supplementary Note~2.

These results demonstrate how an excitonic system within a noisy environment
can exhibit long-lasting coherent features: the observed long-lived
oscillations are the result of coherent interaction of excitonic bands
with an underdamped, quasi-resonant vibration. This vibronic mechanism
requires the vibrational dissipation rate $\gamma_{v}$ to be much
slower than the electronic decoherence rate $\Gamma_{13}$, which
is the case for C8O3, where $\gamma_{v}\lesssim(1{\rm \thinspace ps)^{-1}}$
and $\Gamma_{13}\approx(66\,{\rm fs)}^{-1}$.
The difference in electronic and vibrational decoherence rates can be rationalized from the fact that excitons and vibrations are related to the motion of electrons and nuclei, respectively. The lower mass of electrons as compared to nuclei makes excitons more mobile and therefore more sensitive to environmental fluctuations, such as local electric fields, than vibrations. We note that the vibronic mixing leading to long-lived beating signals in 2D-ES is described by a vibronic coupling that induces coherent energy exchange between excitons and quasi-resonant vibrations (see Supplementary Note~2 for further details):
\begin{equation}
	H_{e-v}=\nu_1\sqrt{S_1}(\ket{3_0}\bra{1_1}+\ket{1_1}\bra{3_0}).\nonumber
\end{equation}
This implies that the vibronic coupling not only induces long-lasting electronic excited-state coherences, but also can mediate population transfer between excitonic bands. In a combination with thermal relaxation of exciton populations, the vibronic coupling may further enhance exciton population transfer and as a result could, in principle, have functional relevance in exciton transport\cite{Womick_JPCB2011,Kolli_JCP2012,Perlik_JCP2015,Killoran_arXiv2015,Schroter_PR2015}.

Interestingly, the different decoherence rates $\Gamma_{g3}\approx2\Gamma_{g1}$
of bands 1 and 3 lead to different amplitudes of the short-lived beating
signals in N11 and R31 (Figs.3\textbf{b} and \textbf{c}), which are
determined by the prefactors $\Gamma_{g1}^{-2}$ and $\Gamma_{g3}^{-1}\Gamma_{g1}^{-1}$,
respectively. The lower decoherence rate of band 1 can be explained
by band 1 being localized on the inner layer, while band 3 is delocalized
over both the inner and outer layers\cite{Didraga_JPCB2004}. As shown
by the response functions for N11 and R31, the overall strength of
the beating signals is proportional to the inverse of the electronic
decoherence rates. It is therefore expected that the beating signal
amplitude would diminish with an increase of the decoherence rate.
This is the case for N22, where the physical situation in terms of
exciton-vibrational resonance ($\Delta\Omega_{32}\approx\nu_{2}\approx470\,{\rm cm}^{-1}$)
is equivalent to N11 ($\Delta\Omega_{31}\approx\nu_{1}\approx668\,{\rm cm}^{-1}$).
The crucial difference is that band 2 has a higher decoherence rate
than band 1, as band 2 is localized on the outer layer exposed to
solvent\cite{Didraga_JPCB2004}. This explains the broader linewidth
of band 2 in absorption and 2D spectra. Using an estimated value of
$\Gamma_{g2}\approx(47\,{\rm fs)}^{-1}$, the presented theory predicts
the strength of N22 to be $5\,\%$ of N11 (see Supplementary Note~2), which is in line
with the experimental observations (Fig.2\textbf{f}). These results
indicate that the experimentally observed long-lived beating signals,
induced by vibronic mixing, require adequately low electronic decoherence
rates, highlighting that resonance between exciton energy splitting
and vibrational frequency alone is not sufficient\cite{Miller_NatureChem2014}.

The presented vibronic model achieves quantitative agreement with the experimental observations. Crucially, the constraints imposed by the observed asymmetric decoherence rates $\Gamma_{g3}\approx2\Gamma_{g1}$ and fast relaxation of exciton population in C8O3 on sub-picosecond timescales\cite{Milota_JPCA2013} rule out incoherent models, where long-lived oscillations are sustained by Markovian correlated fluctuations (see Supplementary Note~3 for a detailed analysis). This further supports our conclusion that the observed experimental data provide evidence for vibronic mixing being the mechanism at play in our system.

We note that our results do not imply that correlated fluctuations can be universally ruled out, as this mechanism could be in place in certain pigment-protein complexes. The notion of correlated fluctuations has been developed for photosynthetic complexes where pigments are embedded in a protein scaffold. The protein has been considered as the potential source of correlated fluctuations in natural light harvesters\cite{Fleming_Science2007,Ishizaki_PCCP2010}. For C8O3, a structural frame such as a protein scaffold is absent and therefore correlated fluctuations are unlikely to induce long-lived oscillatory 2D-signals, which is in line with our observations.

\begin{figure}
\includegraphics{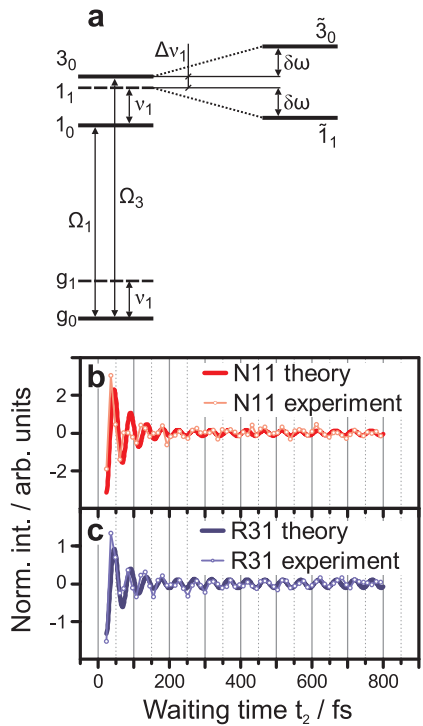} \protect\caption{\textbf{Vibronic model}. \textbf{a}, We consider a vibronic model
for bands 1 and 3 coupled to a vibrational mode with frequency $\nu_{1}\approx668\,{\rm cm}^{-1}$
(see Supplementary Note~2). The vibronic states $\ket{k_{0}}$ and $\ket{k_{1}}$ denote
the vibrational ground and first excited state of an electronic state
$\ket{k}$, respectively, with the single index states $\ket{g}$,
$\ket{1}$ and $\ket{3}$ denoting the electronic ground state and
bands 1 and 3, respectively. The exciton energy splitting $\Delta\Omega_{31}=\Omega_{3}-\Omega_{1}$
between bands 1 and 3 is quasi-resonant with the vibrational frequency
$\nu_{1}$, where the detuning is denoted by $\Delta\nu_{1}=\Delta\Omega_{31}-\nu_{1}$.
The exciton-vibrational coupling between uncoupled states $\ket{3_{0}}$
and $\ket{1_{1}}$ leads to vibronic eigenstates $\mathinner{|\tilde{3}_{0}\rangle}$
and $\mathinner{|\tilde{1}_{1}\rangle}$, each of which is a superposition
of $\ket{3_{0}}$ and $\ket{1_{1}}$, leading to an energy-level shifting
by $\delta\omega$. \textbf{b}, The time trace of N11 where the experimental
results are shown as light red circles, and the theoretical simulation
is shown as a full red line. \textbf{c}, The time trace of R31 where
the experimental results are shown as light blue circles, and the
simulated data are depicted as a full blue line. The root-mean-square
deviation (RMSD) between the experimental results and theoretical
simulation in \textbf{b} and \textbf{c} is 0.92 and 0.59, respectively.}

\label{figure3} 
\end{figure}


\

{\bf Discussion}

We have verified, theoretically and experimentally,
that coherent vibronic coupling in the electronic excited-state manifold
is responsible for the long-lived beating signals observed in 2D spectra
of an artificial light harvester. The relatively simple electronic
and vibrational structure of the investigated molecular aggregate
along with its macroscopic alignment allowed us to rule out the presence
of correlated fluctuations. The specific geometry of our system allowed
us to gain further insights by illustrating the conditions under which
intra-pigment vibrations can prolong electronic coherent effects. The
moderately low decoherence rate of band 1, localized on the inner
layer and protected from solvent, is the basis for exciton-vibrational
coupling as the source of long-lived beating signals. The outer band
2, even though resonantly coupled to a vibration, exhibits a higher
decoherence rate and therefore fails to produce observable oscillations.
We conclude that the mere resonance between excitons and vibrations
does \textit{not} suffice to explain long-lived beating signals. An
adequately low electronic decoherence rate, determined by the interaction
between system and bath, is an equally important prerequisite. 

The influence of vibronic coupling on energy transport in molecular aggregates has been extensively studied in the past, as recently reviewed\cite{Schroter_PR2015}. The vibronic coupling has recently gained new interest (see ref.\cite{Chenu_ARPC2015} for a recent tutorial overview), as it was suggested as a feasible mechanism to explain long-lived oscillations in the 2D spectra
of several natural light harvesting complexes and a photosynthetic
reaction center\cite{Romero_NaturePhys2014,Fuller_NatureChem2014}. 
The requirement of exciton-vibrational resonance is readily satisfied
in such systems, given their numerous excitonic bands and rich vibrational
structures. Incoherent models based upon correlated fluctuations were not ruled out though. Our work provides a quantum mechanical foundation for
enhanced energy transfer based on vibronic coupling. As recently demonstrated,
this mechanism is not limited to natural light harvesting, vibronic
coupling is also of key importance in photovoltaic devices\cite{Falke2014}.


\

{\bf Methods}

{\bf Polarization controlled 2D electronic spectroscopy.} In 2D electronic spectroscopy, three ultrashort
laser pulses generate an optical response of a molecular ensemble,
which is spectrally resolved along both absorption ($\omega_{1}$)
and detection ($\omega_{3}$) frequencies within the laser pulse spectrum.
The absorption frequency $\omega_{1}$ is obtained by precise scanning
of the time delay between the first two pulses and subsequent Fourier
transformation ($t_{1}\rightarrow\omega_{1}$). In detection, the
signal is spectrally dispersed, leading directly to the detection
frequency $\omega_{3}$. Varying time delay $t_{2}$ between pulses
2 and 3 provides information about evolution of the system on a femtosecond
timescale\cite{Brixner_JCP2004,Augulis_OE2011,Augulis_JOSAB2013}.
In order to retrieve the purely absorptive part, the signal induced
by pulses 1-3 is detected in a heterodyned fashion by interfering
it with a phase-stable local oscillator pulse (LO). Polarization control
is achieved by the combination of $\lambda/4$ wave plates and wire
grid polarizers for each of the laser beams to select the desired polarization
with high accuracy. Polarization-resolved 2D experiments change the
relative contributions of distinct pathways depending on the polarization
of the laser pulses, orientation of the transition dipole moments
and isotropy of the sample\cite{Hochstrasser_CP2001}. Rephasing
spectra were acquired with a polarization sequence of $(90^{\circ},0^{\circ},90^{\circ},0^{\circ})$
for pulses (1,\,2,\,3,\,LO), in contrast to non-rephasing spectra,
where the time ordering of the first two pulses is reversed, leading
to a polarization sequence of $(0^{\circ},90^{\circ},90^{\circ},0^{\circ})$. The polarization scheme used for rephasing spectra (Fig.1\textbf{d})
shows $0^{\circ}$ was defined to be parallel to the sample flow direction,
depicted as a white arrow in Fig.1\textbf{a}. For a macroscopically aligned sample, this particular polarization sequence selects pathways stemming from interband coherences and vibronic mixing\cite{Plenio_JCP2013,Jonas_PNAS2012}, discussed throughout the paper, while pathways with all-parallel transition dipole moments such as ground state bleach, stimulated emission, excited state absorption and also vibrational wave packet excitation are suppressed. For the details regarding the experimental methods, see Supplementary Note~1. To subtract
the non-oscillatory signals from 2D spectra, we employed a decay associated
spectra analysis\cite{Milota_JPCA2013}, where the population decays
were fitted by a sum of three 2D-spectra with individual decay constants.
The $\omega_{2}$-maps in Fig.~2 were obtained using Fourier transformation
($t_{2}\rightarrow\omega_{2}$) with zero-padding up to $2^{7}$ data
points. All measurements were carried out at room temperature.


\


\

{\bf Acknowledgements}

The authors would like to thank Valentyn I. Prokhorenko for help in 2D-DAS analysis.
C.N.L. and J.H. acknowledge funding by the Austrian Science Fund (FWF):
START project Y 631-N27 and by COST Action CM1202 - PERSPECT-H2O.
J.L., F.C.-S., S.F.H. and M.B.P. acknowledge funding by the EU STREP PAPETS and QUCHIP, the
ERC Synergy Grant BioQ, the Deutsche Forschungsgemeinschaft (DFG)
within the SFB/TRR21 and an Alexander von Humboldt Professorship.
J.P. acknowledges funding by the Spanish Ministerio de Econom{\'{i}}a
y Competitividad under Project No. FIS2012-30625. D.P. and D.Z. acknowledge
funding by the Swedish Research Council and Knut and Alice Wallenberg
Foundation.


\

{\bf Author contributions}

D.P., D.Z. and J.H. designed and conducted experiments. H.v.B. was
responsible for sample preparation, structural characterization and
Raman measurements. J.L., F.C.-S., C.N.L., D.P., J.P., and J.H. analyzed
the data. J.L., F.C.-S., S.F.H., J.H. and M.B.P. developed theory. All
authors discussed the results and wrote the manuscript.


\

{\bf Competing financial interests}

The authors declare no competing financial interests.


\

{\bf Correspondences} and requests for materials should be addressed to J.H. (juergen.hauer@tuwien.ac.at).


\newpage

\begin{widetext}

\

{\bf\Large Supplementary Information}


\section{Experiment}

\subsection{Sample preparation}

The monomer, tetrachlorobenzimidacarbocyanine chromophore with two attached hydrophobic octyl groups (FEW-Chemicals, Wolfen, Germany) was dissolved in $10^{-2}$ M NaOH solution to achieve a concentration of $10^{-4}$ M. The solution was then stirred in the dark for several hours. Subsequently, polyvinyl alcohol (PVA) of molecular weight $\sim$130000 was added in 1:10 w/w ratio (monomer:PVA) to slow down the formation of aggregate bundles during the storage of dye solutions. Moreover, the adsorbed PVA chains~\cite{Dekany_2001_SI} obviously prevent the reassembly of double-layered into single-layered tubes upon bundling. This effect was observed recently for another derivative~\cite{Eisele_PNAS2014_SI} (C8S3) of the present dye. The individual tubular aggregates degrade in that case into single-layered tubes, which is accompanied by a dramatic change of absorption spectra. Similar effects were not observed for C8O3 when PVA is present. In particular, the aggregate solutions prepared in the described way were stable for approximately ten days when stirred continuously. Without stirring the spectral signature of the double-layered tubes retained even after 12 weeks of storage~\cite{Berlepsch_JPCB2003_SI}. For 2D experiments, we additionally diluted the sample with $10{}^{-2}$ M NaOH to obtain optical density below 0.3 at 598\,nm at a path length of 200\,$\mu {\rm m}$.

A total sample volume of approximately 10\,ml was circulated through the U-shaped wire-guided jet~\cite{Tauber_RSI2003_SI} by a peristaltic pump (Masterflex C/L) with a flow speed optimized for film stability. Solvent evaporated from the recollecting container was refilled every 4 hours during the course of 13 hour measurement.

\subsection{Data acquisition}

Passively stabilized 2D spectroscopy was described in detail elsewhere~\cite{Brixner_JCP2004_SI}. Briefly, a home-built non-collinear optical parametric amplifier (NOPA) seeded by $180\,{\rm fs}$ pulses at $1030\,{\rm nm}$ from PHAROS (Light Conversion Ltd) was tuned to generate $\sim$$16\,{\rm fs}$ pulses ($80\,{\rm nm}$ full width at half maximum) centered at $580\,{\rm nm}$. The NOPA output was split into four pulses and arranged in the so-called boxcar geometry. Waiting time $t_{2}$ was controlled by a mechanical translation stage (PI), whereas coherence time $t_{1}$ was scanned by inserting a pair of fused silica wedges into the first two pulses. All four pulses were focused and overlapped in the sample. The first three generated a third order nonlinear optical response which is emitted in the photon echo phase-matched direction. This signal was heterodyned with an attenuated fourth pulse, called local oscillator (LO). The resulting interference pattern was spectrally resolved and detected by a CCD camera (PIXIS, Princeton Instruments). Most of the scatter was eliminated by the double-frequency lock-in modulation of the first two pulses~\cite{Augulis_OE2011_SI}. The polarization of each pulse was controlled by the combination of $\lambda/4$ wave plates and wire grid polarizers (contrast ratio $> 800$). The accuracy of the polarization angle was estimated to be $\pm1^{\circ}$, where the unwanted signals were typically suppressed by a factor of $\sim$80 for the selected polarization sequence.

To prevent degradation of the sample, the power and repetition rate of the laser were set to $200\,{\rm pJ/pulse}$ and $40\,{\rm kHz}$, respectively. Spectral resolution of $\sim$$35\,{\rm cm}^{-1}$ for the detection frequency $\omega_{3}$ was determined by the grating, the number of CCD pixels and Fourier filtering of the signal during the standard analysis procedure. Coherence time was scanned from $-300\,{\rm fs}$ to $384\,{\rm fs}$ in $1.5\,{\rm fs}$ steps, providing $\sim$$43\,{\rm cm}^{-1}$ spectral resolution of absorption frequency $\omega_{1}$. Waiting time steps of $12\,{\rm fs}$ were sufficient to resolve oscillatory features up to $1350\,{\rm cm}^{-1}$ with $\sim$$35\,{\rm cm}^{-1}$ resolution along $\omega_{2}$.

\subsection{Polarization-controlled 2D-ES}

The strength of 2D signals is determined by the scalar products of molecular transition dipole moments and pulse polarizations. To take advantage of i) the preferential orientation of the J-aggregate (from here on referred to as C8O3) along the flow direction of the jet and ii) mutually perpendicular transition dipole moments of bands 1(2) and 3 of C8O3, we designed a polarization scheme selective for interband coherences. This is similar to the case of an isotropic sample discussed both theoretically~\cite{Hochstrasser_CP2001_SI} and experimentally~\cite{SchlauCohen_NC2012_SI,Westenhoff_JACS2012_SI}. In the presented experiments, the polarization scheme for rephasing signals reads $(90,0,90,0)$ for beams 1-4, respectively. The first and third pulses, polarized orthogonal (90) to the jet's flow direction, interact with bands 1-3. The second pulse, polarized parallel (0) to the jet's flow direction, interacts preferentially with bands 1 and 2, due to the negligible transition dipole moment of band 3 along this direction. The polarization scheme for non-rephasing spectra reads $(0,90,90,0)$, as the ordering of the first two pulses is reversed. These polarization schemes restrict oscillatory signals induced by interband coherences to the lower cross peak in rephasing spectra (R31) and the lower diagonal peak in non-rephasing spectra (N11), as shown in Figures 2 and 3 of the main text. Non-oscillatory 2D signals were subtracted prior to Fourier transformation $t_2\rightarrow \omega_2$.

\begin{figure}[tbp]
	\includegraphics{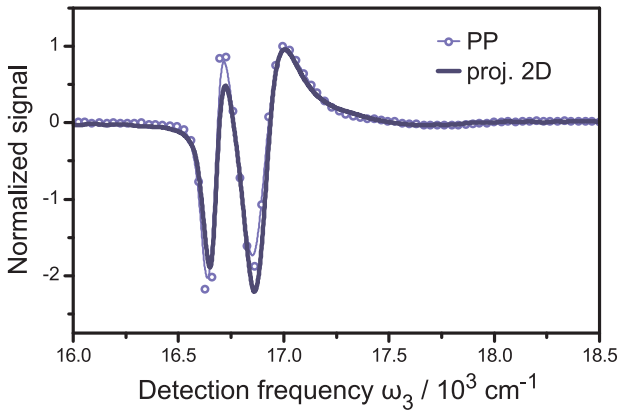}
	\caption{{\bf Pump-probe and projected 2D signal}. Projection of phased polarization-controlled 2D spectra (blue) to all-parallel pump-probe (light blue) at $\mathrm{t_{2}=180\, fs}$.}
	\label{figureS1}
\end{figure}

The polarization-controlled 2D spectra were phased to pump-probe data where pump and probe pulses were polarized in parallel. This procedure is not rigorously correct because in polarization-controlled 2D-ES the first two pulses have different polarization directions while in pump-probe the first two interactions derive from the same pump pulse which naturally means parallel interactions. In other words, the projection slice theorem is strictly speaking not valid for the experiments presented here~\cite{Jonas_ARPC2003_SI}. Despite this discrepancy, one can still satisfactorily phase polarization-controlled 2D spectra to all-parallel pump-probe as shown in Supplementary Figure~\ref{figureS1}. One explanation of this is leakage of the much stronger all-parallel signals through the crossed polarizers, meaning that the all-parallel signal still dominates the non-oscillatory part of the (90,0,90,0) 2D-signal. In this work, we decided to phase polarization-controlled 2D data to parallel pump-probe data. We note that the imperfection in phasing parameters only affects the lineshapes of the real and imaginary part of $\omega{}_{2}$ maps, but preserves their amplitude-maps in both lineshape and magnitude. Hence, the difficulties in phasing polarization-controlled 2D spectra discussed above do not affect the conclusions drawn in the main part of the main text, which were based on $\omega{}_{2}$ amplitude-maps. To this end, we found that arbitrary and large changes of the phasing parameters do not alter $\omega{}_{2}$ amplitude-maps shown in Figure 2 of the main text (results not presented). It is noted that sophisticated phasing techniques based on heterodyned transient grating instead of pump-probe offer a correct method to phase crossed-polarization 2D signals~\cite{Milota_OE2013_SI}.


\section{Theory}


\subsection{A vibronic model for bands 1 and 3 of C8O3}

In the following the vibronic model used to describe bands 1 and 3 of C8O3 and simulate 2D spectra is described. We consider coherent interaction of bands 1 and 3 with the intramolecular vibrational modes of frequency $\hbar\nu_1\approx 668\,{\rm cm}^{-1}$, which is quasi-resonant with the exciton energy splitting between bands 1 and 3. The environmental noise induced by background phonons (a phonon bath) is modeled by a Markovian quantum master equation.

\subsubsection{Hamiltonian}

The electronic Hamiltonian of C8O3 that consists of a network of cyanine dye molecules is described by
\begin{align}
	H_e &=\sum_\alpha\hbar E_\alpha \ket{e_\alpha}\bra{e_\alpha}+\sum_{\alpha\neq\beta}\hbar J_{\alpha\beta}\ket{e_\alpha}\mathinner{\langle{e_\beta}|}\\
	&=\sum_k \hbar\Omega_k \ket{k}\bra{k},
\end{align}
where $\ket{e_\alpha}$ represents the excited state of site $\alpha$ (or molecule $\alpha$), $E_\alpha$ denotes the site energy including electronic and reorganization energies, and $J_{\alpha\beta}$ the electronic coupling between sites $\alpha$ and $\beta$. The diagonalization of the electronic Hamiltonian $H_e$ gives rise to the exciton states $\ket{k}=\sum_{\alpha}\ket{e_\alpha}\mathinner{\langle e_\alpha|k\rangle}$ associated with the exciton energies $\Omega_k$, where bands 1 and 3 are denoted by $\ket{1}$ and $\ket{3}$, respectively: $\mathinner{\langle e_\alpha|k\rangle}\in R$ for $E_{\alpha},J_{\alpha\beta}\in R$.

The vibrational modes with frequency $\hbar\nu_1\approx 668\,{\rm cm}^{-1}$ are described by a set of harmonic oscillators
\begin{equation}
	H_v=\sum_\alpha \hbar\nu_1 a^{\dagger}_\alpha a_\alpha,
\end{equation}
where $a^{\dagger}_\alpha$ and $a_\alpha$ represent the creation and annihilation operators, respectively, of the intramolecular vibrational mode of site $\alpha$. The interaction between vibrations and the electronic excitation of molecules is modeled by
\begin{equation}
	H_{e-v}=\sum_\alpha \hbar\nu_1\sqrt{s_1} \ket{e_\alpha}\bra{e_\alpha}(a^{\dagger}_\alpha+a_\alpha),
\end{equation}
where $s_1$ denotes the Huang-Rhys factor of the vibrational modes. In the exciton basis $\{\ket{k}\}$, the interaction Hamiltonian $H_{e-v}$ is represented by
\begin{equation}
	H_{e-v}=\hbar\nu_1\sqrt{s_1}\sum_{k,l}\ket{k}\bra{l}\sum_\alpha \mathinner{\langle k|e_\alpha\rangle\langle e_\alpha|l\rangle}(a^{\dagger}_\alpha+a_\alpha),
	\label{eq:SI_Hev_total}
\end{equation}
where the diagonal terms ($k=l$) lead to adiabatic surfaces in the electronic excited states, called vibrons, while the non-diagonal terms ($k\ne l$) induce coherent transition between different excitons mediated by exciton-vibrational couplings.

In this work, we are interested in the coherent interaction of bands 1 and 3 with the quasi-resonant vibrational modes of frequency $\nu_1$, which is described by the following cross term $\tilde{H}_{e-v}$ in Eq.~(\ref{eq:SI_Hev_total})
\begin{equation}
	\tilde{H}_{e-v}=\hbar\nu_1\sqrt{S_1}(\ket{1}\bra{3}+\ket{3}\bra{1})(\tilde{a}^{\dagger}_1+\tilde{a}_1),
	\label{eqS:H_ev}
\end{equation}
where $\tilde{a}_1=N\sum_\alpha \mathinner{\langle 1|e_\alpha\rangle\langle e_\alpha|3\rangle}a_\alpha$ describes an effective vibrational mode with frequency $\nu_1$. Here $N$ is introduced to normalize the effective vibrational mode, such that $[\tilde{a}_1,\tilde{a}^{\dagger}_1]=1$, leading to an effective Huang-Rhys factor $S_1=s_1/N^2$. This implies that for a given Huang-Rhys factor $s_1$, the effective Huang-Rhys factor $S_1$ is increased as the spatial overlap $\mathinner{\langle 1|e_\alpha\rangle\langle e_\alpha|3\rangle}$ between excitonic wavefunctions of bands 1 and 3 increases, leading to smaller $N$ and larger $S_1=s_1/N^{2}$. The effective Hamiltonian of bands 1 and 3 coupled to the effective vibrational mode is then described by $\tilde{H}=\tilde{H}_e+\tilde{H}_v+\tilde{H}_{e-v}$, where $\tilde{H}_e = \hbar\Omega_1\ket{1}\bra{1}+\hbar\Omega_3\ket{3}\bra{3}$ and $\tilde{H}_v=\hbar\nu_1 \tilde{a}^{\dagger}_1\tilde{a}_1$. We note that the vibrational energy $\hbar\nu_1\approx 668\,{\rm cm}^{-1}$ is higher than the thermal energy $k_B T\approx 208\,{\rm cm}^{-1}$ at room temperature $T=300\,K$, implying that the thermal state of the vibrational mode is well approximated by its ground state. In addition, when exciton-vibrational couplings are sufficiently small, the light-induced vibrational excitation of overtones is negligible due to the small Franck-Condon factors. This is the case for C8O3, where N11 and R31 in 2D spectra can be well described within a subspace spanned by $\{\ket{g_0},\ket{g_1},\ket{1_0},\ket{1_1},\ket{3_0}\}$. Here, $\ket{k_0}$ and $\ket{k_1}$ denote the vibrational ground and first excited states of an electronic state $\ket{k}$, respectively, {\it i.e.}~$(\tilde{H}_e+\tilde{H}_v)\ket{k_l}=\hbar(\Omega_k+l\nu_1)\ket{k_l}$, where $\ket{g}$ represents the electronic ground state with $\Omega_g=0$. In this scenario, $\{\ket{1_0},\ket{3_0}\}$ can be directly excited by light from the ground state $\ket{g_0}$, while $\{\ket{g_1},\ket{1_1}\}$ has an extremely low transition probability due to small Franck-Condon factors. Nonetheless, $\{\ket{g_1},\ket{1_1}\}$ can be populated through exciton-vibrational coupling $\nu_1\sqrt{S_1}$, leading to transition from $\ket{3_0}$ to $\ket{1_1}$, and subsequently to $\ket{g_1}$ via emission. The coherent transition between $\ket{3_0}$ and $\ket{1_1}$ requires resonance between vibrational frequency $\nu_1$ and exciton energy splitting $\Delta\Omega_{31}=\Omega_3-\Omega_1$ between bands 1 and 3, {\it i.e.}~$\Delta\Omega_{31}\approx\nu_1$.

\subsubsection{Decoherence}

In addition to the coherent interaction of bands 1 and 3 with the effective vibrational mode $\tilde{a}_1$, we consider electronic decoherence induced by background phonons. We characterize the decoherence by two dynamical processes, i) the incoherent population transfer between excitons, called exciton relaxation, and ii) the pure dephasing noise that destroys electronic coherence without exciton population transfer. In addition, we consider iii) relaxation of the effective vibrational mode.

We assume that each cyanine dye molecule is coupled to an independent phonon bath. The Hamiltonian of the background phonons is given by $H_{\rm ph}=\sum_{\xi}\hbar\upsilon_{\xi}b^{\dagger}_{\xi}b_{\xi}$ with the interaction Hamiltonian $H_{e-{\rm ph}}=\sum_{\alpha,\xi}\hbar g_{\alpha\xi}\mathinner{|e_{\alpha}\rangle\langle e_{\alpha}|}(b_{\xi}^{\dagger}+b_{\xi})$ between molecules and phonons, where $b_{\xi}^{\dagger}$ and $b_{\xi}$ denote the creation and annihilation operators, respectively, of a background phonon mode $\xi$. Here $g_{\alpha\xi}$ represents the exciton-phonon coupling between site $\alpha$ and phonon mode $\xi$, which satisfies $g_{\alpha\xi}g_{\beta\xi}=0$ for all $\beta\neq\alpha$, implying that when site $\alpha$ is coupled to the phonon mode $\xi$ with $g_{\alpha\xi}\neq 0$, all the other sites $\beta$ are decoupled from the mode with $g_{\beta\xi}=0$. For the sake of simplicity, we assume that there is no degeneracy in the exciton energies $\Omega_k$, which leads to a relatively simple form of a Markovian quantum master equation. This condition is satisfied even if the exciton energies are close to degeneracy unless they are strictly degenerate, which is satisfied for bands 1 and 3 of our interest. The influence of the background phonons on the vibronic system consisting of bands 1 and 3 with the effective vibrational mode is then described by a Markovian quantum master equation\cite{Breuer_SI}
\begin{equation}
	\frac{d}{dt}\rho(t)=-\frac{i}{\hbar}[\tilde{H},\rho(t)]+{\cal D}_{r}[\rho(t)]+{\cal D}_{d}[\rho(t)]+{\cal D}_{v}[\rho(t)],
	\label{eq:QME_vibronic}
\end{equation}
where $\rho(t)$ denotes the reduced vibronic state, while ${\cal D}_{r}[\rho(t)]$, ${\cal D}_{d}[\rho(t)]$ and ${\cal D}_{v}[\rho(t)]$ describe exciton relaxation, pure dephasing noise and relaxation of the effective vibrational mode, respectively. \\*

{\bf i. Exciton relaxation}\\
Here ${\cal D}_{r}[\rho(t)]$ describes exciton relaxation
\begin{align}
	{\cal D}_{r}[\rho(t)]&=\sum_{\omega\neq 0}\sum_{\alpha}\gamma_{\alpha\alpha}(\omega)\left(A_{\alpha}(\omega)\rho(t)A_{\alpha}^{\dagger}(\omega)-\frac{1}{2}\{A_{\alpha}^{\dagger}(\omega)A_{\alpha}(\omega),\rho(t)\}\right),
	\label{eq:relaxation_vibronic}
\end{align}
with $\Delta\Omega_{kl}=\Omega_k-\Omega_l$ denoting the exciton energy splitting between $\ket{k}$ and $\ket{l}$, $A_{\alpha}(\omega)=\sum_{k,l}\delta(\omega,\Delta\Omega_{kl})\left\langle l|e_{\alpha}\right\rangle\left\langle e_{\alpha}|k\right\rangle\ket{l}\bra{k}$ for $\omega\neq 0$, leading to incoherent transition from $\ket{k}$ to $\ket{l}$, where $\delta(i,j)$ denotes the Kronecker delta defined by $\delta(i,j)=1$ if $i=j$ and $\delta(i,j)=0$ otherwise. In Eq.~(\ref{eq:relaxation_vibronic}), $\gamma_{\alpha\alpha}(\omega)$ is defined by
\begin{equation}
	\gamma_{\alpha\alpha}(\omega)=2\pi{\cal J}_{\alpha}(\omega)(n(\omega)+1),
\end{equation}
with $n(\omega)=(\exp(\hbar\omega/k_B T)-1)^{-1}$ representing the Bose-Einstein distribution function at temperature $T$, ${\cal J}_{\alpha}(\omega)$ is the spectral density of site $\alpha$ defined by ${\cal J}_{\alpha}(\omega)=\sum_{\xi}g_{\alpha\xi}^{2}\delta(\omega-\upsilon_\xi)$ if $\omega\ge 0$ and ${\cal J}_{\alpha}(\omega)=-{\cal J}_{\alpha}(-\omega)$ otherwise. Here $\delta(x)$ represents the Dirac delta function defined by $\delta(x)\rightarrow\infty$ if $x=0$ and $\delta(x)=0$ otherwise with $\int_{-\infty}^{\infty}dx\delta(x)=1$.\\* 

{\bf ii. Pure dephasing noise}\\
${\cal D}_{d}[\rho(t)]$ in Eq.~(\ref{eq:QME_vibronic}) describes the pure dephasing noise
\begin{equation}
	{\cal D}_{d}[\rho(t)]=\sum_{\alpha}\gamma_{\alpha\alpha}(0)\left(A_{\alpha}(0)\rho(t)A_{\alpha}^{\dagger}(0)-\frac{1}{2}\{A_{\alpha}^{\dagger}(0)A_{\alpha}(0),\rho(t)\}\right),
	\label{eq:dephasing_vibronic}
\end{equation}
where $A_{\alpha}(0)=\sum_{k}\abs{\left\langle k|e_{\alpha}\right\rangle}^{2}\ket{k}\bra{k}$ destroys electronic coherence without changing exciton populations defined by $\{{\rm Tr}[\bra{k}\rho(t)\ket{k}]\}$.\\

By substituting electronic coherences $\ket{g}\bra{1}$, $\ket{g}\bra{3}$ and $\ket{1}\bra{3}$ to the dissipators ${\cal D}_{r}[\rho(t)]$ and ${\cal D}_{d}[\rho(t)]$ in Eqs.~(\ref{eq:relaxation_vibronic}) and (\ref{eq:dephasing_vibronic}), one can obtain the following electronic decoherence rates $\Gamma_{g1}$, $\Gamma_{g3}$ and $\Gamma_{13}$ of the coherences $\ket{g}\bra{1}$, $\ket{g}\bra{3}$ and $\ket{1}\bra{3}$
\begin{align}
	\Gamma_{g1}&=\frac{1}{2}\sum_{l\neq 1}\gamma_{1\rightarrow l}+\gamma_{g1},\label{eq:Gg1}\\
	\Gamma_{g3}&=\frac{1}{2}\sum_{l\neq 3}\gamma_{3\rightarrow l}+\gamma_{g3},\label{eq:Gg3}\\
	\Gamma_{13}&=\frac{1}{2}\sum_{l\neq 1}\gamma_{1\rightarrow l}+\frac{1}{2}\sum_{l\neq 3}\gamma_{3\rightarrow l}+\gamma_{13},
\end{align}
where $\gamma_{k\rightarrow l}$ denotes the incoherent population transfer rate from band $k$ to $l$
\begin{equation}
	\gamma_{k\rightarrow l}=\sum_{\alpha}\gamma_{\alpha\alpha}(\Delta\Omega_{kl})\mathinner{|\langle l|e_{\alpha}\rangle\langle e_{\alpha}|k\rangle|}^{2}\ge 0,
\end{equation}
while $\gamma_{g1}$ and $\gamma_{g3}$ represent the pure dephasing rates of the coherences $\ket{g}\bra{1}$ and $\ket{g}\bra{3}$, respectively,
\begin{align}
	\gamma_{g1}&=\frac{1}{2}\sum_{\alpha}\abs{\left\langle 1|e_{\alpha}\right\rangle}^{2}\gamma_{\alpha\alpha}(0)\mathinner{|\langle 1|e_{\alpha}\rangle |}^{2},\label{eq:g1}\\
	\gamma_{g3}&=\frac{1}{2}\sum_{\alpha}\abs{\left\langle 3|e_{\alpha}\right\rangle}^{2}\gamma_{\alpha\alpha}(0)\mathinner{|\langle 3|e_{\alpha}\rangle |}^{2},\label{eq:g3}
\end{align}
and $\gamma_{13}$ represents the pure dephasing rate of the inter-exciton coherence $\ket{1}\bra{3}$ between bands 1 and 3
\begin{equation}
	\gamma_{13}=\frac{1}{2}\sum_{\alpha}\left(\abs{\left\langle 1|e_{\alpha}\right\rangle}^{2}-\abs{\left\langle 3|e_{\alpha}\right\rangle}^{2}\right)\gamma_{\alpha\alpha}(0)\left(\mathinner{|\langle 1|e_{\alpha}\rangle |}^{2}-\mathinner{|\langle 3|e_{\alpha}\rangle |}^{2}\right).
	\label{eq:13}
\end{equation}
These results imply that the inter-exciton dephasing rate $\gamma_{13}$ should be lower than the sum of the other dephasing rates $\gamma_{g1}$ and $\gamma_{g3}$ when there is a spatial overlap between excitonic wavefunctions of bands 1 and 3
\begin{equation}
	\gamma_{13}=\gamma_{g1}+\gamma_{g3}-\sum_{\alpha}\abs{\left\langle 1|e_{\alpha}\right\rangle}^{2}\gamma_{\alpha\alpha}(0)\mathinner{|\langle 3|e_{\alpha}\rangle |}^{2}\le \gamma_{g1}+\gamma_{g3},
\end{equation}
with $\gamma_{\alpha\alpha}(0)\ge 0$ for all $\alpha$, the equality $\gamma_{13}=\gamma_{g1}+\gamma_{g3}$ holds if and only if there is no spatial overlap between excitonic wavefunctions, {\it i.e.}~$\abs{\left\langle 1|e_{\alpha}\right\rangle}^{2}\abs{\left\langle 3|e_{\alpha}\right\rangle}^{2}=0$ for all $\alpha$, or the spectral densities ${\cal J}_{\alpha}(\omega)$ of the molecules shared by bands 1 and 3 do not induce pure dephasing noise by $\gamma_{\alpha\alpha}(0)=0$ for all sites $\alpha$ satisfying $\abs{\left\langle 1|e_{\alpha}\right\rangle}^{2}\abs{\left\langle 3|e_{\alpha}\right\rangle}^{2}\neq 0$. This implies that even if each molecule is coupled to an independent phonon bath, the spatial overlap between excitonic wavefunctions can reduce the inter-exciton dephasing rate $\gamma_{13}$. Here the independent phonon baths of the molecules shared by excitons effectively form a common phonon bath coupled to both excitons, leading to a partial dephasing-free subspace. For instance, if bands 1 and 3 have perfect spatial overlap, {\it i.e.}~$\abs{\left\langle 1|e_{\alpha}\right\rangle}^{2}=\abs{\left\langle 3|e_{\alpha}\right\rangle}^{2}$ for all $\alpha$, while the orthogonality between them is satisfied by the phases of $\left\langle 1|e_{\alpha}\right\rangle\left\langle e_{\alpha}|3\right\rangle$, {\it i.e.}~$\mathinner{\langle 1|3\rangle}=\sum_{\alpha}\mathinner{\langle 1|e_\alpha\rangle\langle e_\alpha |3\rangle}=0$, the inter-exciton dephasing rate $\gamma_{13}$ will become zero, as each $A_{\alpha}(0)=\sum_{k}\abs{\left\langle k|e_{\alpha}\right\rangle}^{2}\ket{k}\bra{k}=\abs{\left\langle 1|e_{\alpha}\right\rangle}^{2}(\ket{1}\bra{1}+\ket{3}\bra{3})+\sum_{k\neq 1,3}\abs{\left\langle k|e_{\alpha}\right\rangle}^{2}\ket{k}\bra{k}$ forms a dephasing-free subspace $\ket{1}\bra{1}+\ket{3}\bra{3}$ of bands 1 and 3. Since band 1 is localized on the inner layer of C8O3, while band 3 is delocalized on both the inner and outer layers\cite{Didraga_JPCB2004_SI}, there is a partial spatial overlap between excitonic wavefunctions, leading to $\gamma_{13}<\gamma_{g1}+\gamma_{g3}$. The spatial overlap is also required for a non-zero value of the effective Huang-Rhys factor $S_1$, which is responsible for the long-lived beating signals observed in the experiment, as will be discussed later.

In addition, the inter-exciton dephasing rate $\gamma_{13}$ has a non-zero lower bound when the dephasing rates $\gamma_{g1}$ and $\gamma_{g3}$ are different in magnitude. The dephasing rates in Eqs.~(\ref{eq:g1})-(\ref{eq:13}) can be expressed as $\gamma_{g1}=\mathinner{|\vec{v}_{1}|}^{2}$, $\gamma_{g3}=\mathinner{|\vec{v}_{3}|}^{2}$ and $\gamma_{13}=\mathinner{|\vec{v}_{1}-\vec{v}_{3}|}^{2}$ with the real vectors $\vec{v}_k$ defined by $\vec{v}_k=2^{-1/2}\hat{\gamma}^{1/2}\vec{w}_k$, where $\vec{w}_k$ is a real vector with elements $\abs{\left\langle k|e_{\alpha}\right\rangle}^{2}\ge 0$ representing the delocalization of an exciton state $\ket{k}$ in the site basis $\{\ket{e_\alpha}\}$, while $\hat{\gamma}$ is a diagonalized matrix with elements $\gamma_{\alpha\alpha}(0)\ge 0$, leading to a positive matrix $\hat{\gamma}^{1/2}$ defined by $\hat{\gamma}=\hat{\gamma}^{1/2}\hat{\gamma}^{1/2}$. From the triangle inequality, $\mathinner{|\vec{v}_{1}-\vec{v}_{3}|}+\mathinner{|\vec{v}_{3}|}\ge\mathinner{|\vec{v}_{1}|}$ and $\mathinner{|\vec{v}_{1}-\vec{v}_{3}|}+\mathinner{|\vec{v}_{1}|}\ge\mathinner{|\vec{v}_{3}|}$, the inter-exciton dephasing rate $\gamma_{13}$ is bounded from below by $(\sqrt{\gamma_{g1}}-\sqrt{\gamma_{g3}}~)^{2}$, leading to $(\sqrt{\gamma_{g1}}-\sqrt{\gamma_{g3}}~)^{2}\le \gamma_{13}<\gamma_{g1}+\gamma_{g3}$. Therefore, the electronic decoherence rate $\Gamma_{13}$ of the inter-exciton coherence $\ket{1}\bra{3}$ is constrained by
\begin{equation}
	\frac{1}{2}\sum_{l\neq 1}\gamma_{1\rightarrow l}+\frac{1}{2}\sum_{l\neq 3}\gamma_{3\rightarrow l}+\left[\left(\Gamma_{g1}-\frac{1}{2}\sum_{l\neq 1}\gamma_{1\rightarrow l}\right)^{1/2}-\left(\Gamma_{g3}-\frac{1}{2}\sum_{l\neq 3}\gamma_{3\rightarrow l}\right)^{1/2}~\right]^{2}\le\Gamma_{13}<\Gamma_{g1}+\Gamma_{g3},
	\label{eq:gamma_13_range}
\end{equation}
with $\gamma_{gk}=\Gamma_{gk}-\frac{1}{2}\sum_{l\neq k}\gamma_{k\rightarrow l}$ from Eqs.~(\ref{eq:Gg1}) and (\ref{eq:Gg3}). Here the population transfer rates ($\gamma_{1\rightarrow l}$ and $\gamma_{3\rightarrow l}$) and electronic decoherence rates ($\Gamma_{g1}$ and $\Gamma_{g3}$) can be estimated using experimentally measured 2D spectra, which will be discussed later.\\*

{\bf iii. Relaxation of quasi-resonant vibrations}\\
Finally, ${\cal D}_{v}[\rho(t)]$ in Eq.~(\ref{eq:QME_vibronic}) describes the relaxation of the effective vibrational mode
\begin{equation}
	{\cal D}_{v}[\rho(t)]=\gamma_{v}(n(\nu_1)+1)\left(2\tilde{a}_{1}\rho(t) \tilde{a}_{1}^{\dagger}-\{\tilde{a}_{1}^{\dagger}\tilde{a}_{1},\rho(t)\}\right)+\gamma_{v}n(\nu_1)\left(2\tilde{a}_{1}^{\dagger}\rho(t) \tilde{a}_{1}-\{\tilde{a}_{1}\tilde{a}_{1}^{\dagger},\rho(t)\}\right).
	\label{eqS:thermalization}
\end{equation}
Since $n(\nu_1)\approx 0.04$ at room temperature $T=300\,K$ due to the high vibrational energy $\hbar\nu_1\gg k_{B}T$, Eq.~(\ref{eqS:thermalization}) can be reduced to
\begin{equation}
	{\cal D}_{v}[\rho(t)]\approx \gamma_{v}\left(2\tilde{a}_{1}\rho(t) \tilde{a}_{1}^{\dagger}-\{\tilde{a}_{1}^{\dagger}\tilde{a}_{1},\rho(t)\}\right),
	\label{eqS:vib_dissipation}
\end{equation}
which describes the dissipation of the vibrational mode with the rate of $\gamma_{v}$.


\begin{figure}[tbp]
	\includegraphics{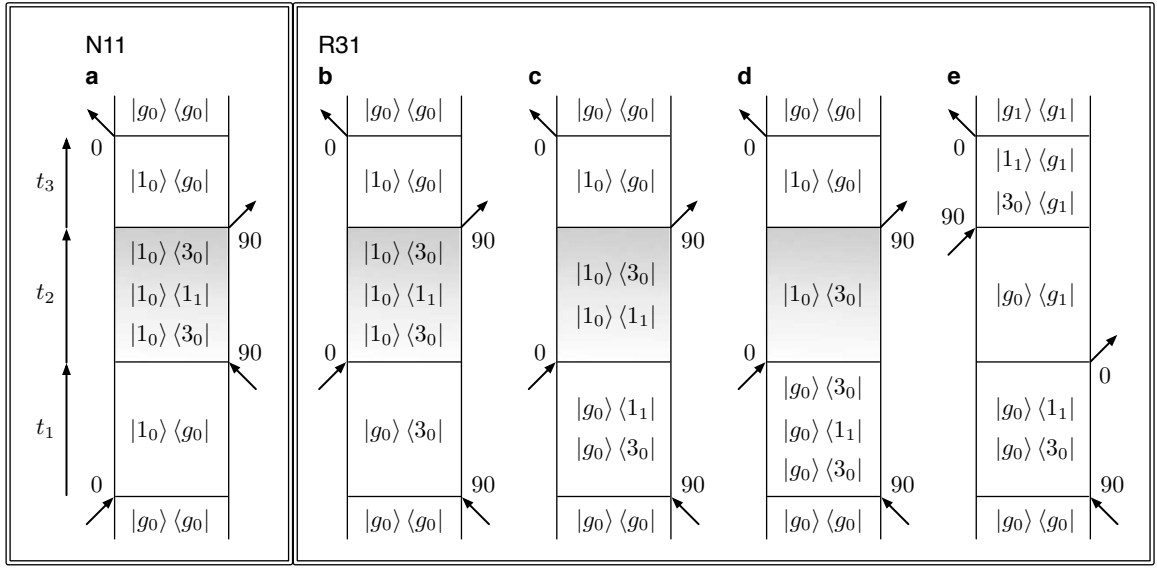}
	\caption{{\bf Feynman diagrams contributing to the beating signals in N11 and R31 represented in uncoupled state basis}. {\bf a}, The stimulated emission diagram contributing to the beating signals in N11. Here time runs upwards and the electronic transitions induced by light are denoted by arrows: 0 and 90 denote the polarization of light, parallel and normal to the longitudinal axis of C8O3, respectively ({\it cf}.~Figure 1 in the main text). The time interval between pulses is called coherence time $t_1$, waiting time $t_2$, and rephasing time $t_3$ for the first and second, the second and third, and the third excitation pulse and the emerging signal, respectively. The Fourier transform along $t_1$ and $t_3$ leads to the absorption and detection frequencies denoted by $\omega_1$ and $\omega_3$, respectively. {\bf b-e}, The stimulated emission and ground state bleaching diagrams contributing to the beating signals in R31. In {\bf a-d}, grey shaded waiting time periods during $t_2$ highlight vibronic coherences in the electronic excited states. In {\bf e}, on the other hand, the vibronic system is in the electronic ground state during $t_2$.}
	\label{figureS2}
\end{figure}


\subsubsection{The response function for N11}

Here we derive the response function for the beating signals in N11, which is a diagonal peak in non-rephasing spectra centered at $(\omega_1,\omega_3)\approx(\Omega_1,\Omega_1)$.

In Fig.~\ref{figureS2}a, the Feynman diagram contributing to the beating signals in N11 after the employed $(0,90,90,0)$ excitation is displayed. As the thermal state of the effective vibrational mode at room temperature is well approximated by its ground state ($\hbar\nu_1\gg k_{B}T$), the initial state of the vibronic system is given by $\ket{g_0}\bra{g_0}$. After excitation to $\ket{1_0}\bra{g_0}$ by the first pulse, the dynamics of $\ket{1_0}\bra{g_0}$ during coherence time $t_1$ is governed by a time evolution super-operator ${\cal U}(t_1)$ determined by the quantum master equation in Eq.~(\ref{eq:QME_vibronic})
\begin{equation}
	{\cal U}(t_1)\ket{1_0}\bra{g_0}=e^{(-i\Omega_{1}-\Gamma_{g1})t_1}\ket{1_0}\bra{g_0},
\end{equation}
for which the Fourier transform is given by
\begin{equation}
	\int_{0}^{\infty}dt_1 e^{i\omega_1 t_1}{\cal U}(t_1)\ket{1_0}\bra{g_0}=-\frac{1}{i(\omega_1-\Omega_1)-\Gamma_{g1}}\ket{1_0}\bra{g_0},
\end{equation}
where the prefactor $-(i(\omega_1-\Omega_1)-\Gamma_{g1})^{-1}$ determines the lineshape of N11 along the $\omega_1$-axis, which is centered at $\omega_1=\Omega_1$ with a linewidth of $2\Gamma_{g1}$. By the second pulse, $\ket{1_0}\bra{g_0}$ becomes $\ket{1_0}\bra{3_0}$, which evolves during waiting time $t_2$ into a mixture of $\ket{1_0}\bra{3_0}$ and $\ket{1_0}\bra{1_1}$, mediated by exciton-vibrational coupling, scaling with $\nu_1\sqrt{S_1}$. The time evolution of $\ket{1_0}\bra{3_0}$ is formally expressed as
\begin{equation}
	{\cal U}(t_2)\ket{1_0}\bra{3_0}=\ket{1_0}\bra{3_0}e^{K t_2}=f(t_2)\ket{1_0}\bra{3_0}+g(t_2)\ket{1_0}\bra{1_1},
	\label{eq:A(t2)}
\end{equation}
where $K$ is a non-Hermitian operator describing both the Hamiltonian dynamics and decoherence
\begin{align}
	K&=(i\Delta\Omega_{31}-\Gamma_{13})\ket{3_0}\bra{3_0}+(i\nu_1-\gamma_{v})\ket{1_1}\bra{1_1}+i\nu_1\sqrt{S_1}(\ket{3_0}\bra{1_1}+\ket{1_1}\bra{3_0}).
	\label{eq:K}
\end{align}
Here we evaluate $f(t_2)$ in Eq.~(\ref{eq:A(t2)}), which describes the case that $\ket{1_0}\bra{3_0}$ becomes $\ket{1_0}\bra{g_0}$ by the third pulse, as shown in Fig.~\ref{figureS2}a. By diagonalizing the non-Hermitian operator $K$, one can show that $f(t_2)$ is given by
\begin{equation}
	f(t_2)=\sum_{k=1}^{2}\frac{1}{2}\left(1+(-1)^k\frac{x-y}{\sqrt{(x-y)^2+4z^2}}\right)\exp\left[\frac{1}{2}\left(x+y+(-1)^k\sqrt{(x-y)^2+4z^2}\right)t_2\right],
\end{equation}
where $x=i\Delta\Omega_{31}-\Gamma_{13}$, $y=i\nu_1-\gamma_{v}$ and $z=i\nu_1\sqrt{S_1}$. Finally, $\ket{1_0}\bra{g_0}$ evolves during rephasing time $t_3$
\begin{equation}
	{\cal U}(t_3)\ket{1_0}\bra{g_0}=e^{(-i\Omega_{1}-\Gamma_{g1})t_3}\ket{1_0}\bra{g_0},
\end{equation}
for which the Fourier transform leads to the lineshape $-(i(\omega_3-\Omega_1)-\Gamma_{g1})^{-1}$ of N11 along the $\omega_3$-axis. Therefore, the response function for N11 is given by
\begin{equation}
	R_{1g}(\omega_1,t_2,\omega_3)=\mu_{1p}^{2}\mu_{3n}^{2}\frac{1}{i(\omega_1-\Omega_1)-\Gamma_{g1}}\frac{1}{i(\omega_3-\Omega_1)-\Gamma_{g1}}f(t_2),
	\label{eq:N11}
\end{equation}
where $\mu_{1p}$ denotes the transition dipole moment of band 1 for light polarized parallel to the longitudinal axis of C8O3, while $\mu_{3n}$ represents the transition dipole moment of band 3 for light polarized normal to the axis. This is due to the $(0,90,90,0)$ polarization scheme employed for measuring non-rephasing spectra in the experiment, as schematically shown in Fig.~\ref{figureS2}a. It is notable that all the Feynman diagrams in Figs.~\ref{figureS2}a-e can be induced by $(90,90,90,90)$ excitation where all the pulses are polarized normal to the longitudinal axis: band 1 can be excited or de-excited by both 0 and 90 polarizations, although with higher efficiency for light polarized at 0. For $(90,90,90,90)$ excitation, the overall dipole strength $\mu_{1p}^{2}\mu_{3n}^{2}$ in Eq.~(\ref{eq:N11}) is decreased to $\mu_{1n}^{2}\mu_{3n}^{2}$ with $\mu_{1p}^{2}>\mu_{1n}^{2}$, as band 1 is mainly polarized along the longitudinal axis of C8O3, as shown in the linear dichroism spectrum in Figure 1 of the main text. This implies that the $(0,90,90,0)$ polarization scheme for non-rephasing spectra enhances the signal-to-noise ratio when compared to the $(90,90,90,90)$ excitation. Similarly, the signal-to-noise ratio of rephasing spectra is enhanced by $(90,0,90,0)$ excitation.

The lineshape function $(i(\omega_1-\Omega_1)-\Gamma_{g1})^{-1}(i(\omega_3-\Omega_1)-\Gamma_{g1})^{-1}$ in Eq.~(\ref{eq:N11}) shows that N11 is centered at $(\omega_1,\omega_3)=(\Omega_1,\Omega_1)$ with a symmetric linewidth $2\Gamma_{g1}$ along $\omega_1$- and $\omega_3$-axes. When $(\omega_1,\omega_3)=(\Omega_1,\Omega_1)$, the lineshape function is reduced to $\Gamma_{g1}^{-2}$, implying that the amplitude of the N11 peak is proportional to $\Gamma_{g1}^{-2}$, which is decreased as the linewidth $2\Gamma_{g1}$ increases. The time-dependent term $f(t_2)$ in Eq.~(\ref{eq:N11}) describes the evolution of N11 during waiting time $t_2$. In the absence of the exciton-vibrational coupling ($S_1=0$), $f(t_2)$ is reduced to
\begin{equation}
	f(t_2)|_{S_1=0}=e^{(i\Delta\Omega_{31}-\Gamma_{13})t_2},
	\label{eq:A(t2)S0}
\end{equation}
implying that the coherence $\ket{1_0}\bra{3_0}$ oscillates with the frequency of the exciton energy splitting $\Delta\Omega_{31}$ and decays with the electronic decoherence rate $\Gamma_{13}$. Conversely, in the presence of the exciton-vibrational coupling ($S_1>0$), $f(t_2)$ is expressed as
\begin{equation}
	f(t_2)=\frac{1}{2}\left(1+\frac{x-y}{\sqrt{(x-y)^2+4z^2}}\right)e^{[i(\Delta\Omega_{31}+\delta\omega)-\Gamma_{13}+\delta\gamma]t_2}+\frac{1}{2}\left(1-\frac{x-y}{\sqrt{(x-y)^2+4z^2}}\right)e^{[i(\nu_1-\delta\omega)-\gamma_{v}-\delta\gamma]t_2},
	\label{eq:exact_A(t2)}
\end{equation}
where $i\delta\omega+\delta\gamma=2^{-1}[\sqrt{(x-y)^2 +4z^2}-(x-y)]$, which satisfies $\delta\omega>0$ and $\delta\gamma>0$ for $\Delta\Omega_{31}>\nu_1$ and $\Gamma_{13}>\gamma_v$, which is the case for C8O3. There are several notable features that result from the vibronic coupling evident in Eq.~(\ref{eq:exact_A(t2)}). i) The first term, proportional to $e^{[i(\Delta\Omega_{31}+\delta\omega)-\Gamma_{13}+\delta\gamma]t_2}$, oscillates with a frequency of $\Delta\Omega_{31}'=\Delta\Omega_{31}+\delta\omega$, which is higher than the exciton energy splitting $\Delta\Omega_{31}$, and decays with the rate of $\Gamma_{13}-\delta\gamma$, which is lower than the electronic decoherence rate $\Gamma_{13}$ shown in Eq.~(\ref{eq:A(t2)S0}). These are the characteristics of the vibronic coherence $\ket{1_0}\mathinner{\langle\tilde{3}_{0}|}$, where $\mathinner{\langle{\tilde{3}_0}|}$ is one of the left eigenstates of $K$ in the form of $\mathinner{\langle{\tilde{3}_0}|}\propto \bra{3_0}+\xi\bra{1_1}$ with $|\xi|<1$. The vibronic eigenstate $\mathinner{\langle{\tilde{3}_0}|}$ has a higher energy-level than $\bra{3_0}$ due to the exciton-vibrational coupling, leading to $\Delta\Omega_{31}'>\Delta\Omega_{31}$ (see Figure 3a in the main text). Additionally, the amplitude of $\ket{1_0}\mathinner{\langle{\tilde{3}_0}|}$ in $\ket{1_0}\bra{1_1}$ denoted by $\xi$ leads to a longer lifetime than the coherence $\ket{1_0}\bra{3_0}$ that has no vibrational character, or in other words, the lifetime borrowing effect. ii) Conversely, the second term in Eq.~(\ref{eq:exact_A(t2)}),  proportional to $e^{[i(\nu_1-\delta\omega)-\gamma_{v}-\delta\gamma]t_2}$, exhibits characteristics of the other vibronic coherence $\ket{1_0}\mathinner{\langle{\tilde{1}_1}|}$, where $\mathinner{\langle{\tilde{1}_1}|}\propto \bra{1_1}-\xi\bra{3_0}$ is the other left eigenstate of $K$. The second term oscillates with frequency $\nu_1'=\nu_1-\delta\omega$, which is lower than the vibrational frequency $\nu_1$ due to the exciton-vibrational coupling (see Figure 3a in the main text). It also decays with the rate of $\gamma_{v}+\delta\gamma$, which is higher than the vibrational decoherence rate $\gamma_{v}$ of $\ket{1_0}\bra{1_1}$ due to the amplitude of $\ket{1_0}\mathinner{\langle{\tilde{1}_1}|}$ in $\ket{1_0}\bra{3_0}$ denoted by $\xi$. iii) We add that the vibronic states $\mathinner{\langle{\tilde{3}_0}|}\propto \bra{3_0}+\xi\bra{1_1}$ and $\mathinner{\langle{\tilde{1}_1}|}\propto \bra{1_1}-\xi\bra{3_0}$ are the eigenstates of the non-Hermitian operator $K$ in Eq.~(\ref{eq:K}) describing both Hamiltonian dynamics and decoherence, where $\xi$ depends on the parameters of the Hamiltonian as well as decoherence rates. These states are different from the eigenstates of the Hamiltonian $\tilde{H}$, which do not depend on decoherence rates, and their difference becomes non-negligible when the electronic decoherence rate $\Gamma_{13}$ is comparable to or larger than the exciton-vibrational coupling $\nu_1\sqrt{S_1}$, as is the case for C8O3.


\begin{figure}[tbp]
	\includegraphics{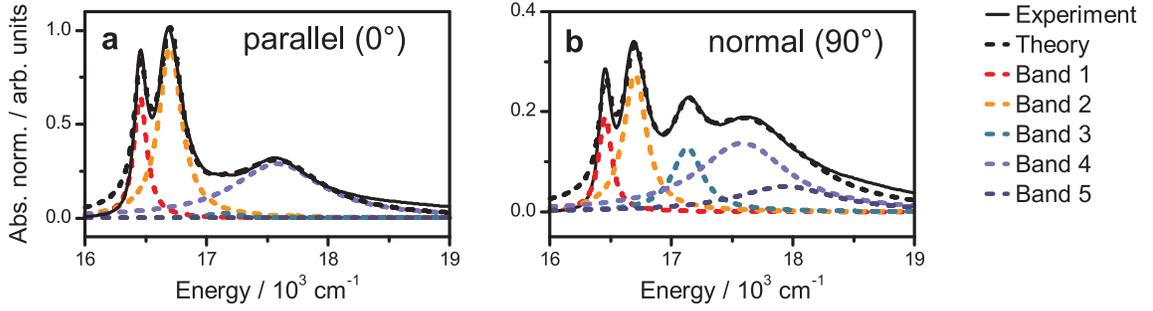}
	\caption{{\bf Absorption spectrum of C8O3}. {\bf a}, Absorption spectrum with light polarized parallel to the longitudinal axis of C8O3. Experimental and theoretical results are shown as a black solid line and a black dashed line, respectively. Theoretical results were modeled by a sum of Lorentzian functions, which describe bands 1-5 of C8O3. Each Lorentzian function is shown as a colored dashed line. {\bf b}, Absorption spectrum with light polarized normal to the longitudinal axis of C8O3. Note that the vertical scales in {\bf a} and {\bf b} are different.}
	\label{figureS3}
\end{figure}

By fitting experimental 2D spectra to the theoretical prediction of N11 and R31, which will be discussed later, we found that $\hbar\Delta\Omega_{31}\approx 720\,{\rm cm^{-1}}$, $\hbar\nu_1\approx 668\,{\rm cm^{-1}}$, $\hbar\Gamma_{g1}\approx 65\,{\rm cm^{-1}}$, $\hbar\Gamma_{g3}\approx 150\,{\rm cm^{-1}}$, $\hbar\Gamma_{13}\approx 80\,{\rm cm}^{-1}$, $S_1=0.0006$ ({\it cf}.~$\hbar\nu_1\sqrt{S_1}\approx 16\,{\rm cm}^{-1}$) and $\gamma_v \lesssim (1\,{\rm ps})^{-1}$. The estimated electronic decoherence rates $\Gamma_{g1}$ and $\Gamma_{g3}$ reproduce well the absorption spectrum of C8O3, as shown in Fig.~\ref{figureS3}, where experimental and theoretical results are shown as a black solid line and a black dashed line, respectively. The theoretical results were modeled by a sum of the Lorentzian functions with linewidths $2\Gamma_{gk}$ for $k\in\{1,2,3,4,5\}$, each of which describes the absorption of band $k$: each Lorentzian function is shown as a colored dashed line. The estimated values of the parameters lead to $\hbar\delta\omega\approx 1.6\,{\rm cm^{-1}}$ and $\hbar\delta\gamma\approx 2.1\,{\rm cm^{-1}}$, which are smaller than the experimental resolution of $\sim$$\,40\,{\rm cm}^{-1}$. This implies that for the case of C8O3, we can approximate $\Delta\Omega_{31}'$ and $\nu_1'$ by $\Delta\Omega_{31}$ and $\nu_1$, respectively, with $\delta\omega\approx 0$ and $\delta\gamma\approx 0$. More specifically, when the exciton-vibrational coupling is sufficiently small, such that $\nu_1\sqrt{S_1}<\abs{i\Delta\nu_{1}-\Gamma_{13}}$ with $\Delta\nu_{1}=\Delta\Omega_{31}-\nu_1$, and the dissipation rate of the vibrational mode is negligible within the timescale of the total measurement time, {\it i.e.}~$\gamma_{v}\approx 0$, the response function determining N11 in Eq.~(\ref{eq:N11}) is reduced to
\begin{equation}
	R_{1g}(\omega_1,t_2,\omega_3)\approx\mu_{1p}^{2}\mu_{3n}^{2}\frac{1}{i(\omega_1-\Omega_1)-\Gamma_{g1}}\frac{1}{i(\omega_3-\Omega_1)-\Gamma_{g1}}\left[e^{(i\Delta\Omega_{31}-\Gamma_{13})t_2}(1-\epsilon_{2}^{2})+e^{i\nu_1 t_2}\epsilon_{2}^{2}\right],
	\label{eq:R1g}
\end{equation}
with $\epsilon_{2}$ representing the degree of vibronic mixing during waiting time $t_2$
\begin{equation}
	\epsilon_{2}=i\nu_1\sqrt{S_1}(i\Delta\nu_{1}-\Gamma_{13})^{-1},
	\label{eq:vibronic_mixing}
\end{equation}
where the vibronic eigenstates $\mathinner{\langle{\tilde{3}_0}|}$ and $\mathinner{\langle{\tilde{1}_1}|}$ are approximated by $\mathinner{\langle{\tilde{3}_0}|}\propto\bra{3_0}+\epsilon_{2}\bra{1_1}$ and $\mathinner{\langle{\tilde{1}_1}|}\propto\bra{1_1}-\epsilon_{2}\bra{3_0}$, respectively, with $|\epsilon_{2}|^{2}\ll1$ (in the main text, $\epsilon_2$ was denoted by $\epsilon$ for the sake of simplicity). It can be seen in Eq.~(\ref{eq:vibronic_mixing}) that $|\epsilon_{2}|$ increases as the exciton-vibrational coupling $\nu_1\sqrt{S_1}$ increases or the detuning $|\Delta\nu_{1}|=|\Delta\Omega_{31}-\nu_1|$ between exciton splitting and vibrational frequency decreases. This implies that the vibronic mixing of the coherences $\ket{1_0}\bra{3_0}$ and $\ket{1_0}\bra{1_1}$ requires resonance between excitons and vibrations and induces the observed long-lived beating signal in N11. In this respect, when $|\epsilon_{2}|$ decreases as a result of a high electronic decoherence rate $\Gamma_{13}$, the coherence $\ket{1_0}\bra{3_0}$ generated by the second pulse (see Fig.~\ref{figureS2}a) will decohere too quickly and thereby suppressing the vibronic mixing of $\ket{1_0}\bra{3_0}$ and $\ket{1_0}\bra{1_1}$ during waiting time $t_2$, which in turn will suppress the long-lived beating signal in N11. This is related to the fact that $\epsilon_{2}$ is proportional to the exciton-vibrational coupling $\nu_1\sqrt{S_1}$ and the amplitude of the long-lived component $e^{i\nu_1 t_2}$ in Eq.~(\ref{eq:R1g}) is proportional to $\epsilon_{2}^2$. As such, when $\nu_1\sqrt{S_1}<\abs{i\Delta\nu_{1}-\Gamma_{13}}$, the response function for N11 can be effectively described by two transitions between $\ket{1_0}\bra{3_0}$ and $\ket{1_0}\bra{1_1}$ during waiting time $t_2$, mediated by exciton-vibrational coupling $\nu_1\sqrt{S_1}$, {\it i.e.}~$\ket{1_0}\bra{3_0}\rightarrow\ket{1_0}\bra{1_1}\rightarrow\ket{1_0}\bra{3_0}$, within the timescale of the electronic decoherence rate $\Gamma_{13}$, as shown in Fig.~\ref{figureS2}a. When the condition of $\nu_1\sqrt{S_1}<\abs{i\Delta\nu_{1}-\Gamma_{13}}$ is not satisfied, the response function for N11 is represented by $R_{1g}(\omega_1,t_2,\omega_3)=\sum_{n=0}^{\infty}h_{n}(\omega_1,t_2,\omega_3)(i\nu_1\sqrt{S_1})^{2n}$ with the higher order terms proportional to $(i\nu_1\sqrt{S_1})^{2n}$, which describe multiple transitions between $\ket{1_0}\bra{3_0}$ and $\ket{1_0}\bra{1_1}$ during $t_2$.

In summary, when $\nu_1\sqrt{S_1}<\abs{i\Delta\nu_{1}-\Gamma_{13}}$ and $\gamma_{v}\approx 0$, the response function for N11 at $(\omega_1,\omega_3)=(\Omega_1,\Omega_1)$ is given by
\begin{equation}
	R_{1g}(t_2)\approx\mu_{1p}^{2}\mu_{3n}^{2}\Gamma_{g1}^{-2}\left[e^{(i\Delta\Omega_{31}-\Gamma_{13})t_2}+e^{i\nu_1 t_2}\epsilon_{2}^{2}\right],
	\label{eq:N11summary}
\end{equation}
with $\epsilon_{2}$ defined in Eq.~(\ref{eq:vibronic_mixing}). The lineshape of N11 is symmetric along $\omega_1$- and $\omega_3$-axes with a linewidth of $2\Gamma_{g1}$. These results are in line with the experimental observations shown in Figures 2 and 3 of the main text.


\subsubsection{The response function for R31}

Here we provide the response function for the beating signals in R31, which is the cross peak in the rephasing spectra centered at $(\omega_1,\omega_3)\approx(\Omega_3,\Omega_1)$. The response function for R31 can be derived using the same approach described above for N11. Here we provide the results without derivation. Figs.~\ref{figureS2}b-e show the Feynman diagrams contributing to the beating signals in R31. In Figs.~\ref{figureS2}b-d, the vibronic system is in the electronic excited states during $t_2$, while in Fig.~\ref{figureS2}e, the system is in the electronic ground state, each of which is called the stimulated emission (SE) and ground state bleaching (GSB) diagram, respectively.

When $\nu_1\sqrt{S_1}<\abs{i\Delta\nu_{1}-\Gamma_{13}}$, $\nu_1\sqrt{S_1}<\abs{i\Delta\nu_{1}+\Gamma_{g1}-\Gamma_{g3}}$ and $\gamma_{v}\approx 0$, which are satisfied for the case of C8O3, the contribution of the SE diagrams to R31 is approximated by
\begin{align}
	R_{2g}(\omega_1,t_2,\omega_3)
	&\approx\mu_{1p}^{2}\mu_{3n}^{2}\left\{\frac{1}{-i(\omega_1-\Omega_3)-\Gamma_{g3}}\frac{1}{i(\omega_3-\Omega_1)-\Gamma_{g1}}\left[e^{(i\Delta\Omega_{31}-\Gamma_{13})t_2}(1-\epsilon_{2}^2)+e^{i\nu_1 t_2}\epsilon_{2}^2\right]\right.\label{eq:R2g}\\
	&\quad+\left(-\frac{1}{-i(\omega_1-\Omega_3)-\Gamma_{g3}}+\frac{1}{-i(\omega_1-\Omega_3+\Delta\nu_{1})-\Gamma_{g1}}\right)\frac{1}{i(\omega_3-\Omega_1)-\Gamma_{g1}}\left(-e^{(i\Delta\Omega_{31}-\Gamma_{13})t_2}+e^{i\nu_1 t_2}\right)\epsilon_{1}\epsilon_{2}\nonumber\\
	&\quad\left.+\left(-\frac{1}{-i(\omega_1-\Omega_3)-\Gamma_{g3}}+\frac{1}{-i(\omega_1-\Omega_3+\Delta\nu_{1})-\Gamma_{g1}}\right)\frac{1}{i(\omega_3-\Omega_1)-\Gamma_{g1}}e^{(i\Delta\Omega_{31}-\Gamma_{13})t_2}\epsilon_{1}^2\nonumber\right\},
\end{align}
with $\epsilon_{1}$ representing the degree of vibronic mixing during coherence time $t_1$
\begin{equation}
	\epsilon_{1}=i\nu_1\sqrt{S_1}(i\Delta\nu_{1}+\Gamma_{g1}-\Gamma_{g3})^{-1},
	\label{eq:vibronic_mixing_1}
\end{equation}
where $\Delta\Omega_{31}'$ and $\nu_1'$ are approximated by $\Delta\Omega_{31}$ and $\nu_1$, respectively. More specifically, $\epsilon_{1}$ is associated with the transition between $\ket{g_0}\bra{3_0}$ and $\ket{g_0}\bra{1_1}$ during $t_1$, while $\epsilon_{2}$ is associated with the transition between $\ket{1_0}\bra{3_0}$ and $\ket{1_0}\bra{1_1}$ during $t_2$. In Eq.~(\ref{eq:R2g}), the first term proportional to $\epsilon_{2}^2$ describes the transition $\ket{1_0}\bra{3_0}\rightarrow\ket{1_0}\bra{1_1}\rightarrow\ket{1_0}\bra{3_0}$ during $t_2$ (see Fig.~\ref{figureS2}b), the second term proportional to $\epsilon_{1}\epsilon_{2}$ describes the transition $\ket{g_0}\bra{3_0}\rightarrow\ket{g_0}\bra{1_1}$ during $t_1$ and the subsequent transition $\ket{1_0}\bra{1_1}\rightarrow\ket{1_0}\bra{3_0}$ during $t_2$ (see Fig.~\ref{figureS2}c), and the last term proportional to $\epsilon_{1}^2$ describes the transition $\ket{g_0}\bra{3_0}\rightarrow\ket{g_0}\bra{1_1}\rightarrow\ket{g_0}\bra{3_0}$ during $t_1$ (see Fig.~\ref{figureS2}d). In the second and last terms, the lineshape function along the $\omega_1$-axis contains $(-i(\omega_1-\Omega_3+\Delta\nu_{1})-\Gamma_{g1})^{-1}$, which describes the presence of a sub-peak centered at $\omega_1=\Omega_3-\Delta\nu_{1}=\Omega_1+\nu_1<\Omega_3$ with a linewidth of $2\Gamma_{g1}$, which is induced by exciton-vibrational coupling. However, due to the condition of $|\epsilon_{1}|^{2}\ll 1$ and $|\epsilon_{2}|^{2}\ll 1$, the first term in Eq.~(\ref{eq:R2g}) determines the overall lineshape of R31, which is given by $(-i(\omega_1-\Omega_3)-\Gamma_{g3})^{-1}(i(\omega_3-\Omega_1)-\Gamma_{g1})^{-1}$ that is centered at $(\omega_1,\omega_3)=(\Omega_3,\Omega_1)$ with the asymmetric linewidths of $2\Gamma_{g3}$ and $2\Gamma_{g1}$ along $\omega_1$- and $\omega_3$-axes, respectively.

The contribution of the GSB diagram to R31, with ground state coherence~\cite{Jonas_PNAS2012_SI} during $t_2$, is given by
\begin{align}
	R_{3g}(\omega_1,t_2,\omega_3)\approx\left(\frac{1}{-i(\omega_1-\Omega_3)-\Gamma_{g3}}-\frac{1}{-i(\omega_1-\Omega_3+\Delta\nu_{1})-\Gamma_{g1}}\right)\left(\frac{1}{i(\omega_3-\Omega_1-\Delta\nu_{1})-\Gamma_{g3}}-\frac{1}{i(\omega_3-\Omega_1)-\Gamma_{g1}}\right)e^{i\nu_1 t_2}\epsilon_{1}\epsilon_{3},
	\label{eq:R3g}
\end{align}
with $\epsilon_{3}$ representing the vibronic mixing during $t_3$
\begin{equation}
	\epsilon_{3}=-i\nu_1\sqrt{S_1}(-i\Delta\nu_{1}+\Gamma_{g1}-\Gamma_{g3})^{-1},
	\label{eq:vibronic_mixing_3}
\end{equation}
which is associated with the transition $\ket{3_0}\bra{g_1}\rightarrow\ket{1_1}\bra{g_1}$ during $t_3$ shown in Fig.~\ref{figureS2}e. Here the vibrational frequency $\nu_1$ in $e^{i\nu_1 t_2}$ stems from the vibrational coherence $\ket{g_0}\bra{g_1}$ in the electronic ground-state manifold and not the result of the approximation $\delta\omega\approx 0$.


\begin{figure}[tbp]
	\includegraphics{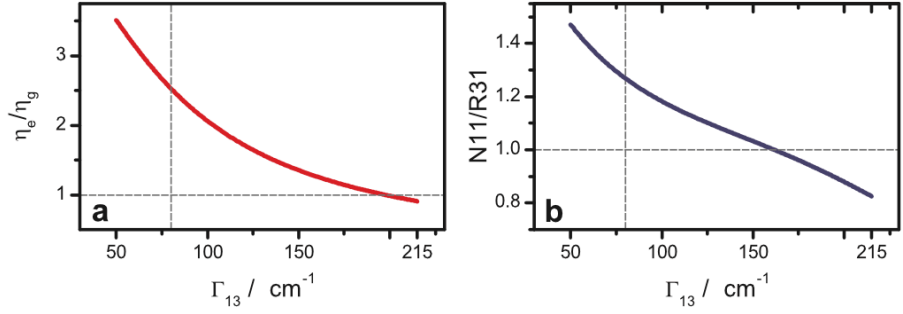}
	\caption{{\bf Long-lived beating signals in N11 and R31}. {\bf a}, The ratio $|\eta_e/\eta_g|$ between the contributions of the vibronic and vibrational coherences to the long-lived beating signal in R31. For the experimentally estimated value of $\Gamma_{13}$, marked by a vertical dashed line, the contribution of the vibronic coherence is greater than the vibrational coherence. {\bf b}, The ratio $\Gamma_{g3}(\Gamma_{g1}|\eta_e-\eta_g|)^{-1}$ between the amplitudes of the long-lived beating signals in N11 ($\propto\Gamma_{g1}^{-2}$) and R31 ($\propto\Gamma_{g3}^{-1}\Gamma_{g1}^{-1}|\eta_e-\eta_g|$). In both {\bf a} and {\bf b}, we take the values of the parameters estimated from experimental results. According to Eq.~(\ref{eq:gamma_13_range}), $\hbar(\Gamma_{g1} +\Gamma_{g3})\approx 215\,{\rm cm}^{-1}$ is the theoretical upper bound for $\Gamma_{13}$.}
	\label{figureS4}
\end{figure}

In summary, when $\nu_1\sqrt{S_1}<\abs{i\Delta\nu_{1}-\Gamma_{13}}$, $\nu_1\sqrt{S_1}<\mathinner{|i\Delta\nu_{1}+\Gamma_{g1}-\Gamma_{g3}|}$ and $\gamma_{v}\approx 0$, the response function for R31 at $(\omega_1,\omega_3)=(\Omega_3,\Omega_1)$ is given by
\begin{align}
	R_{2g}(t_2)+R_{3g}(t_2)\approx\mu_{1p}^2\mu_{3n}^2\Gamma_{g3}^{-1}\Gamma_{g1}^{-1}[e^{(i\Delta\Omega_{31}-\Gamma_{13})t_2}+e^{i\nu_1 t_2}\epsilon_{2}^2 (\eta_e-\eta_g)],
	\label{eq:R31summary}
\end{align}
where $\eta_e=(\Gamma_{g1}+\Gamma_{13})(\Gamma_{g1}+i\Delta\nu_{1})^{-1}$ stems from the SE diagrams shown in Figs.~\ref{figureS2}b-d, while $\eta_g=(\Gamma_{13}-i\Delta\nu_{1})^{2}(\Gamma_{g1}+i\Delta\nu_{1})^{-1}(\Gamma_{g3}+i\Delta\nu_{1})^{-1}$ originates from the GSB diagram shown in Fig.~\ref{figureS2}e. It is interesting to note that the origin of the long-lived oscillations at R31, whether predominantly vibrational or vibronic, depends upon the electronic decoherence rates $\{\Gamma_{g1},\Gamma_{g3},\Gamma_{13}\}$ and detuning $\Delta\nu_{1}=\Delta\Omega_{31}-\nu_1$. In Fig.~\ref{figureS4}a, the ratio $|\eta_e / \eta_g|$ between the contributions of the vibronic and vibrational coherences to the long-lived beating signal in R31 is displayed as a function of the inter-exciton decoherence rate $\Gamma_{13}$, where $\{\Gamma_{g1},\Gamma_{g3},\Delta\nu_{1}\}$ are taken to be the values estimated from experimental results. Here $|\eta_e / \eta_g|>1$ implies that the long-lived beating signal in R31 is dominated by the vibronic coherence $\ket{1}\mathinner{\langle\tilde{1}_1|}$ in the electronic excited-state manifold. By fitting the experimentally measured beating signals in N11 and R31 to the theoretical model, we found that $\hbar\Gamma_{13}\approx 80\,{\rm cm}^{-1}$, which is marked by a vertical dashed line in Fig.~\ref{figureS4}a, where the contribution of the vibronic coherence is $\sim$$\,2.5$ times greater than the vibrational coherence. These results imply that the long-lived beating signal in R31 is dominated by vibronic coherence, originating from electronic excited states. It is notable that the vibronic contribution outweighs the vibrational part for a wide range of $\Gamma_{13}$. This is mainly due to the fact that the vibronic mixing $\epsilon_{2}\propto(i\Delta\nu_1-\Gamma_{13})^{-1}$ during $t_2$ depends on the inter-exciton decoherence rate $\Gamma_{13}$, while the other vibronic mixings $\epsilon_{1}\propto(i\Delta\nu_1+\Gamma_{g1}-\Gamma_{g3})^{-1}$ and $\epsilon_{3}\propto(-i\Delta\nu_1+\Gamma_{g1}-\Gamma_{g3})^{-1}$ during $t_1$ and $t_3$ are independent of $\Gamma_{13}$. Considering that vibronic coherence depends on $\epsilon_{2}$ (see Eq.~(\ref{eq:R2g}) and Figs.~\ref{figureS2}b and c), while vibrational coherence depends on $\epsilon_{1}\epsilon_{3}$ (see Eq.~(\ref{eq:R3g}) and Fig.~\ref{figureS2}e), the vibronic contribution is increased as $\Gamma_{13}$ decreases. We note that these results are in line with the experimental observation that the amplitude of the long-lived beating signal in N11 is greater than that of R31 (see Figures 3b and c in the main text). In Fig.~\ref{figureS4}b, the ratio $\Gamma_{g3}(\Gamma_{g1}|\eta_e-\eta_g|)^{-1}$ between the amplitudes of the long-lived beating signals in N11 and R31 is displayed as a function of the inter-exciton decoherence rate $\Gamma_{13}$. Here the amplitude of the long-lived beating signal in N11 is greater than R31, {\it i.e.}~$\Gamma_{g3}(\Gamma_{g1}|\eta_e-\eta_g|)^{-1}>1$, for a range of $\Gamma_{13}$ where the vibronic coherence dominates the long-lived beating signal in R31, as shown in Fig.~\ref{figureS4}a.


\subsubsection{Numerical simulation of N11 and R31}

So far the analytic form of the response functions for N11 and R31 were derived with the assumption that the vibronic system is well described within the subspace of the vibrational ground and first excited states, which is valid for a small Huang-Rhys factor $S_1$. To clarify the validity of this assumption, we performed numerical simulation of the beating signals in N11 and R31 with higher vibrational excited states, {\it i.e.}~$\{\ket{g_0},\ket{g_1},\cdots,\ket{g_n},\ket{1_0},\ket{1_1},\cdots,\ket{1_n},\ket{3_0},\ket{3_1},\cdots,\ket{3_n}\}$ with $n\ge1$. We found that the theoretical beating signals converge for $n\ge 1$ and the numerical results are well matched to the analytical results. Here the electronic decoherence was modeled by a convex combination of two effective dissipators, {\it i.e.}~$p{\cal D}_{1}[\rho(t)]+(1-p){\cal D}_{2}[\rho(t)]$ with $0\le p\le 1$, where the dissipators are given by
\begin{align}
	{\cal D}_{1}[\rho(t)]&=\Gamma_{g1}(2\ket{1}\bra{1}\rho(t)\ket{1}\bra{1}-\{\ket{1}\bra{1},\rho(t)\})+\Gamma_{g3}(2\ket{3}\bra{3}\rho(t)\ket{3}\bra{3}-\{\ket{3}\bra{3},\rho(t)\}),\\
	{\cal D}_{2}[\rho(t)]&=2\left(\sqrt{\Gamma_{g1}}\ket{1}\bra{1}+\sqrt{\Gamma_{g3}}\ket{3}\bra{3}\,\right)\rho(t)\left(\sqrt{\Gamma_{g1}}\ket{1}\bra{1}+\sqrt{\Gamma_{g3}}\ket{3}\bra{3}\,\right)-\{\Gamma_{g1}\ket{1}\bra{1}+\Gamma_{g3}\ket{3}\bra{3},\rho(t)\}.
\end{align}
By substituting electronic coherences $\ket{g}\bra{1}$ and $\ket{g}\bra{3}$ to the dissipators, one can show that both ${\cal D}_{1}[\rho(t)]$ and ${\cal D}_{2}[\rho(t)]$ give rise to the same set of decoherence rates $\Gamma_{g1}$ and $\Gamma_{g3}$ for $\ket{g}\bra{1}$ and $\ket{g}\bra{3}$, respectively, implying that the decoherence rates of $\ket{g}\bra{1}$ and $\ket{g}\bra{3}$ are independent of the value of $p$ in the convex combination. For $\ket{1}\bra{3}$, on the other hand, ${\cal D}_{1}[\rho(t)]$ and ${\cal D}_{2}[\rho(t)]$ lead to different decoherence rates $\Gamma_{g1}+\Gamma_{g3}$ and $(\sqrt{\Gamma_{g1}}-\sqrt{\Gamma_{g3}}~)^{2}$, respectively. This enables us to vary the inter-exciton decoherence rate $\Gamma_{13}$ within a range of $(\sqrt{\Gamma_{g1}}-\sqrt{\Gamma_{g3}}~)^{2}\le\Gamma_{13}\le\Gamma_{g1}+\Gamma_{g3}$ by changing the value of $p$ in the convex combination. In addition to the electronic decoherence, the relaxation of the vibrational mode was modeled by Eq.~(\ref{eqS:thermalization}) in the simulations. We found that Eq.~(\ref{eqS:thermalization}) can be approximated by Eq.~(\ref{eqS:vib_dissipation}) due to the high vibrational frequency ($\hbar\nu_1\gg k_B T$).


\begin{figure}[tbp]
	\includegraphics{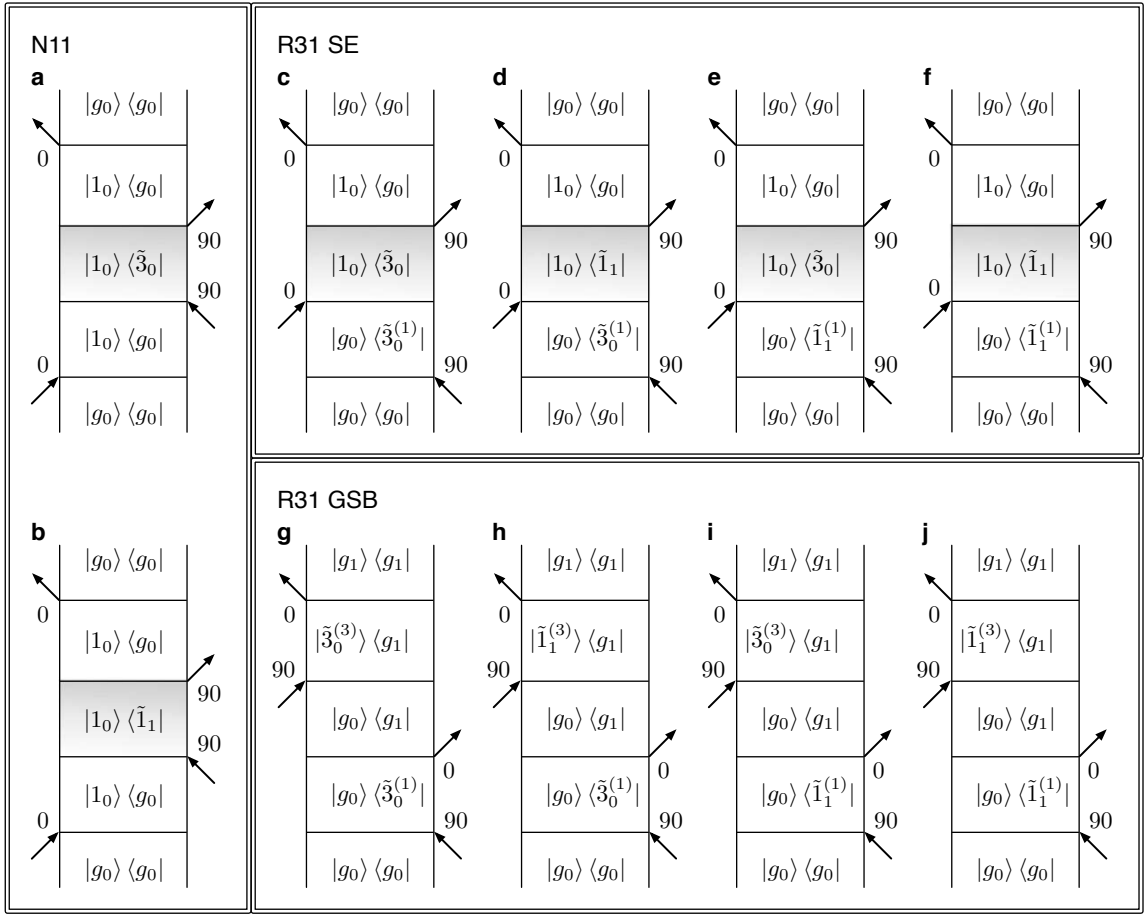}
	\caption{{\bf Feynman diagrams contributing to the beating signals in N11 and R31 represented in vibronic eigenbasis}. {\bf a,b}, The stimulated emission diagrams contributing to the beating signals in N11. {\bf c-f}, The stimulated emission diagrams contributing to the beating signals in R31. {\bf g-j}, The ground state bleaching diagrams contributing to the beating signals in R31.}
	\label{figureS5}
\end{figure}


\subsubsection{Feynman diagrams represented in vibronic eigenbasis}

Here we provide the Feynman diagrams for N11 and R31 represented in the vibronic eigenbasis of the time evolution super-operator ${\cal U}(t)$, which are equivalent to the Feynman diagrams in the uncoupled state basis shown in Fig.~\ref{figureS2}.

For N11, the vibronic mixing $\epsilon_2$ takes place during waiting time $t_2$ ({\it cf}.~Fig.~\ref{figureS2}a), where the vibronic coherences responsible for the short-lived and long-lived beating signals in N11 are given by
\begin{align}
	\ket{1_0}\mathinner{\langle\tilde{3}_0|}&=(1+\epsilon_{2}^{2})^{-1/2}(\ket{1_0}\bra{3_0}+\epsilon_{2}\ket{1_0}\bra{1_1}),\label{eq:coht21}\\
	\ket{1_0}\mathinner{\langle\tilde{1}_1|}&=(1+\epsilon_{2}^{2})^{-1/2}(\ket{1_0}\bra{1_1}-\epsilon_{2}\ket{1_0}\bra{3_0}),\label{eq:coht22}
\end{align}
respectively, where the vibronic eigenstates $\mathinner{\langle\tilde{3}_0|}\propto\bra{3_0}+\epsilon_{2}\bra{1_1}$ and $\mathinner{\langle\tilde{1}_1|}\propto\bra{1_1}-\epsilon_{2}\bra{3_0}$ are normalized by $(1+\epsilon_{2}^{2})^{-1/2}$, not by $(1+|\epsilon_{2}|^{2})^{-1/2}$, due to the biorthogonality of the eigenstates of the non-Hermitian operator $K$ in Eq.~(\ref{eq:K}). When the light-induced vibrational excitation of overtones, {\it i.e.}~$\bra{1_1}$, is negligible due to the small Franck-Condon factors, the transition dipole moments of the vibronic eigenstates $\mathinner{\langle\tilde{3}_0|}$ and $\mathinner{\langle\tilde{1}_1|}$ are determined by their amplitudes in $\bra{3_0}$, each of which is given by $\mu_{3n}(1+\epsilon_{2}^{2})^{-1/2}$ and $-\mu_{3n}(1+\epsilon_{2}^{2})^{-1/2}\epsilon_{2}$, respectively. Here $\mu_{3n}$ denotes the transition dipole moment of $\bra{3_0}$. In the eigenbasis, the Feynman diagrams responsible for the short-lived and long-lived beating signals in N11 are described by Figs.~\ref{figureS5}a and b, respectively. Given that there are two transitions between $\bra{g_0}$ and $\mathinner{\langle\tilde{3}_0|}$ (and also between $\bra{g_0}$ and $\mathinner{\langle\tilde{1}_1|}$) by the second and third pulses, the square of the transition dipole moments of $\mathinner{\langle\tilde{3}_0|}$ and $\mathinner{\langle\tilde{1}_1|}$ is reflected in the response function, each of which is given by $\mu_{3n}^{2}(1+\epsilon_{2}^{2})^{-1}\approx\mu_{3n}^{2}(1-\epsilon_{2}^{2})$ and $\mu_{3n}^{2}\epsilon_{2}^{2}$, respectively. This is in line with the analytic form of the response function for N11 shown in Eq.~(\ref{eq:R1g}).

For R31, on the other hand, vibronic mixing takes place during coherence, waiting and rephasing times ($t_1$, $t_2$, $t_3$, respectively, {\it cf.}~Figs.~\ref{figureS2}b-e). The vibronic mixing $\epsilon_1$ during coherence time $t_1$ leads to the vibronic eigenstates $\mathinner{\langle\tilde{3}_{0}^{(1)}|}\propto\bra{3_0}+\epsilon_{1}\bra{1_1}$ and $\mathinner{\langle\tilde{1}_{1}^{(1)}|}\propto\bra{1_1}-\epsilon_{1}\bra{3_0}$, where vibronic coherences during $t_1$ are represented by
\begin{align}
	\ket{g_0}\mathinner{\langle\tilde{3}_{0}^{(1)}|}&=(1+\epsilon_{1}^{2})^{-1/2}(\ket{g_0}\bra{3_0}+\epsilon_{1}\ket{g_0}\bra{1_1}),\label{eq:coht11}\\
	\ket{g_0}\mathinner{\langle\tilde{1}_{1}^{(1)}|}&=(1+\epsilon_{1}^{2})^{-1/2}(\ket{g_0}\bra{1_1}-\epsilon_{1}\ket{g_0}\bra{3_0}).\label{eq:coht12}
\end{align}
Here the superindex $(1)$ of $\mathinner{\langle\tilde{3}_{0}^{(1)}|}$ and $\mathinner{\langle\tilde{1}_{1}^{(1)}|}$ reminds us that the vibronic mixing takes place during coherence time $t_1$: throughout this work, the vibronic eigenstates $\mathinner{\langle\tilde{3}_{0}^{(2)}|}$ and $\mathinner{\langle\tilde{1}_{1}^{(2)}|}$ responsible for the vibronic mixing $\epsilon_2$ during waiting time $t_2$ have, for the sake of simplicity, been denoted by $\mathinner{\langle\tilde{3}_{0}|}$ and $\mathinner{\langle\tilde{1}_{1}|}$, respectively. We note that $\epsilon_1$ in Eq.~(\ref{eq:vibronic_mixing_1}) is different from $\epsilon_2$ in Eq.~(\ref{eq:vibronic_mixing}), as the time evolution of the coherences $\ket{g_0}\bra{3_0}$ and $\ket{g_0}\bra{1_1}$ during coherence time $t_1$ is governed by a different non-Hermitian operator $K_1$
\begin{equation}
	K_1=(i\Omega_3-\Gamma_{g3})\ket{3_0}\bra{3_0}+(i\Omega_1+i\nu_1-\Gamma_{g1}-\gamma_{v})\ket{1_1}\bra{1_1}+i\nu_1\sqrt{S_1}(\ket{3_0}\bra{1_1}+\ket{1_1}\bra{3_0}),
\end{equation}
defined by ${\cal U}(t_1)\ket{g_0}\bra{3_0}=\ket{g_0}\bra{3_0}e^{K_1 t_1}$. In the eigenbasis, the SE diagrams shown in Figs.~\ref{figureS2}b-d can be represented by four diagrams shown in Figs.~\ref{figureS5}c-f, where the transition dipole moments of $\mathinner{\langle\tilde{3}_{0}^{(1)}|}$ and $\mathinner{\langle\tilde{1}_{1}^{(1)}|}$ are given by $\mu_{3n}(1+\epsilon_{1}^{2})^{-1/2}$ and $-\mu_{3n}(1+\epsilon_{1}^{2})^{-1/2}\epsilon_{1}$, respectively. It is notable that the vibronic eigenstates $\mathinner{\langle\tilde{3}_{0}^{(1)}|}$ and $\mathinner{\langle\tilde{1}_{1}^{(1)}|}$ during coherence time $t_1$ are different from the vibronic eigenstates $\mathinner{\langle\tilde{3}_{0}|}$ and $\mathinner{\langle\tilde{1}_{1}|}$ during waiting time $t_2$, as the vibronic system is in a superposition between electronic ground and excited states (see Eqs.~(\ref{eq:coht11}) and (\ref{eq:coht12})) and in the electronic excited-state manifold (see Eqs.~(\ref{eq:coht21}) and (\ref{eq:coht22})), respectively, which leads in general to different values of the vibronic mixings $\epsilon_1$ and $\epsilon_2$. The diagrams shown in Figs.~\ref{figureS5}c-f describe the fact that the vibronic eigenstates $\mathinner{\langle\tilde{3}_{0}^{(1)}|}$ and $\mathinner{\langle\tilde{1}_{1}^{(1)}|}$ can be represented by superpositions of $\mathinner{\langle\tilde{3}_{0}|}$ and $\mathinner{\langle\tilde{1}_{1}|}$. In Figs.~\ref{figureS5}c and d, for instance, the vibronic eigenstate $\mathinner{\langle\tilde{3}_{0}^{(1)}|}$ induced by the first pulse can be represented by a superposition of $\mathinner{\langle\tilde{3}_{0}|}$ and $\mathinner{\langle\tilde{1}_{1}|}$
\begin{align}
	\mathinner{\langle\tilde{3}_{0}^{(1)}|}&=(1+\epsilon_{1}^{2})^{-1/2}(\bra{3_0}+\epsilon_1\bra{1_1})\\
	&=(1+\epsilon_{1}^{2})^{-1/2}(1+\epsilon_{2}^{2})^{-1/2}[(1+\epsilon_{1}\epsilon_{2})\mathinner{\langle\tilde{3}_{0}|}+(\epsilon_{1}-\epsilon_{2})\mathinner{\langle\tilde{1}_{1}|})].
\end{align}
Here the prefactors of $\mathinner{\langle\tilde{3}_{0}|}$ and $\mathinner{\langle\tilde{1}_{1}|}$, {\it i.e.}~$(1+\epsilon_{1}^{2})^{-1/2}(1+\epsilon_{2}^{2})^{-1/2}(1+\epsilon_{1}\epsilon_{2})$ and $(1+\epsilon_{1}^{2})^{-1/2}(1+\epsilon_{2}^{2})^{-1/2}(\epsilon_{1}-\epsilon_{2})$, enable us to introduce two separated diagrams shown in Figs.~\ref{figureS5}c and d, where the prefactors are multiplied to the response function, similar to the transition dipole moment. Similarly, the other vibronic eigenstate $\mathinner{\langle\tilde{1}_{1}^{(1)}|}$ can be represented by a superposition of $\mathinner{\langle\tilde{3}_{0}|}$ and $\mathinner{\langle\tilde{1}_{1}|}$, leading to the prefactors for the diagrams shown in Figs.~\ref{figureS5}e and f. Using the transition dipole moments of $\mathinner{\langle\tilde{3}_{0}|}$ and $\mathinner{\langle\tilde{1}_{1}|}$ induced by the third pulse, one can show that the response function induced by the SE diagrams is given by Eq.~(\ref{eq:R2g}): here the lineshape functions $(-i(\omega_1-\Omega_3)-\Gamma_{g3})^{-1}$ and $(-i(\omega_1-\Omega_3+\Delta\nu_{1})-\Gamma_{g1})^{-1}$ along the $\omega_1$-axis correspond to the diagrams where the vibronic system is in $\ket{g_0}\mathinner{\langle\tilde{3}_{0}^{(1)}|}$ ({\it cf}.~Figs.~\ref{figureS5}c and d) and in $\ket{g_0}\mathinner{\langle\tilde{1}_{1}^{(1)}|}$ ({\it cf}.~Figs.~\ref{figureS5}e and f), respectively, during coherence time $t_1$.

The vibronic mixing $\epsilon_3$ during rephasing time $t_3$ leads to the vibronic eigenstates $\mathinner{|\tilde{3}_{0}^{(3)}\rangle}\propto\ket{3_0}+\epsilon_{3}\ket{1_1}$ and $\mathinner{|\tilde{1}_{1}^{(3)}\rangle}\propto\ket{1_1}-\epsilon_{3}\ket{3_0}$, where vibronic coherences during $t_3$ are represented by
\begin{align}
	\mathinner{|\tilde{3}_{0}^{(3)}\rangle}\bra{g_1}&=(1+\epsilon_{3}^{2})^{-1/2}(\ket{3_0}\bra{g_1}+\epsilon_{3}\ket{1_1}\bra{g_1}),\\
	\mathinner{|\tilde{1}_{1}^{(3)}\rangle}\bra{g_1}&=(1+\epsilon_{3}^{2})^{-1/2}(\ket{1_1}\bra{g_1}-\epsilon_{3}\ket{3_0}\bra{g_1}).
\end{align}
The time evolution of the coherences $\ket{3_0}\bra{g_1}$ and $\ket{1_1}\bra{g_1}$ is governed by a non-Hermitian operator $K_3$
\begin{equation}
	K_3=(-i\Omega_3+i\nu_1-\Gamma_{g3}-\gamma_{v})\ket{3_0}\bra{3_0}+(-i\Omega_1-\Gamma_{g1})\ket{1_1}\bra{1_1}-i\nu_1\sqrt{S_1}(\ket{3_0}\bra{1_1}+\ket{1_1}\bra{3_0}),
\end{equation}
defined by ${\cal U}(t_3)\ket{3_0}\bra{g_1}=e^{K_3 t_3}\ket{3_0}\bra{g_1}$. Similar to the SE diagrams, the GSB diagram shown in Fig.~\ref{figureS2}e can be represented by four diagrams shown in Figs.~\ref{figureS5}g-j. Using the transition dipole moments of the vibronic eigenstates, one can show that the response function induced by the GSB diagrams is given by Eq.~(\ref{eq:R3g}), where the lineshape functions $(i(\omega_3-\Omega_1-\Delta\nu_{1})-\Gamma_{g3})^{-1}$ and $(i(\omega_3-\Omega_1)-\Gamma_{g1})^{-1}$ along the $\omega_3$-axis correspond to the diagrams where the vibronic system is in $\mathinner{|\tilde{3}_{0}^{(3)}\rangle}\bra{g_1}$ and in $\mathinner{|\tilde{1}_{1}^{(3)}\rangle}\bra{g_1}$, respectively, during $t_3$.

These results imply that the Feynman diagrams for N11 and R31 can be represented in both uncoupled state basis and vibronic eigenbasis equivalently, and the analytic form of the response functions in Eqs.~(\ref{eq:R1g}), (\ref{eq:R2g}) and (\ref{eq:R3g}) is independent of the basis chosen to represent the Feynman diagrams.


\begin{figure}[tbp]
	\includegraphics{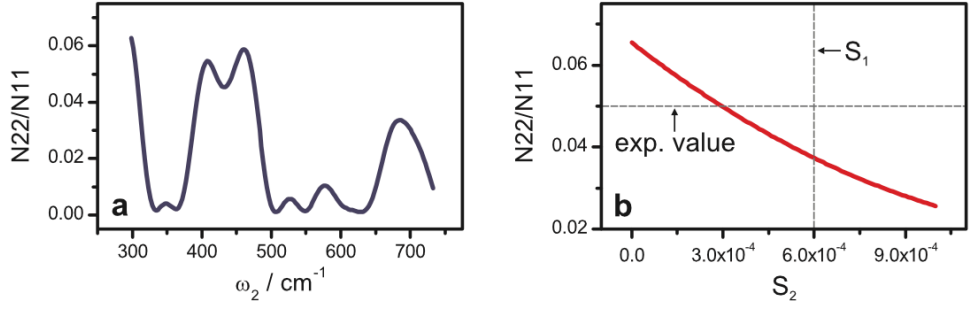}
	\caption{{\bf The relative amplitude of N22 and N11}. {\bf a}, The absolute square of the Fourier transform of the beating signal in N22 as a function of $\omega_2$, which is normalized to the amplitude of N11 at $\hbar\omega_2\approx 705\,{\rm cm}^{-1}$. {\bf b}, Theoretical results of the ratio between N22 and N11 are displayed as a function of the Huang-Rhys factor $S_2$. The Huang-Rhys factor $S_1=0.0006$ of the vibrational mode with frequency $\hbar\nu_1\approx 668\,{\rm cm}^{-1}$ is marked by a vertical dashed line.}
	\label{figureS6}
\end{figure}


\subsection{The response function for N22}

Here we provide a vibronic model for bands 2 and 3 of C8O3, where bands 2 and 3 are coupled to the intramolecular vibrational modes with frequency $\hbar\nu_2\approx 470\,{\rm cm}^{-1}$.

In Fig.~\ref{figureS6}a, the absolute square of the Fourier transform of the beating signal in N22 is displayed as a function of $\omega_2$, which is normalized by the amplitude of N11 at $\hbar\omega_2\approx 705\,{\rm cm}^{-1}$. The amplitude of N22 is maximized around $\hbar\omega_2\approx 460\,{\rm cm}^{-1}$ with an amplitude in the range of $5\,\%$ of the N11 peak. When bands 2 and 3 are coupled to a vibrational mode with frequency $\nu_2$ mediated by an effective Huang-Rhys factor $S_2$, the response function for N22 is given by
\begin{equation}
	R_{1g}(t_2)\approx\mu_{2p}^{2}\mu_{3n}^{2}\Gamma_{g2}^{-2}\left[e^{(i\Delta\Omega_{32}-\Gamma_{23})t_2}+e^{i\nu_2 t_2}\left(\frac{i\nu_{2}\sqrt{S_2}}{i\Delta\nu_2-\Gamma_{23}}\right)^{2}\right],
\end{equation}
with $\Delta\nu_2=\Delta\Omega_{32}-\nu_2$, $\mu_{2p}$ denotes the transition dipole moment of band 2 for light polarized parallel to the longitudinal axis of C8O3 and $\hbar\Gamma_{g2}\approx 110\,{\rm cm}^{-1}$ represents the electronic decoherence rate of band 2, both of which can be estimated using the absorption spectrum shown in Fig.~\ref{figureS3}. From the experimentally measured beating signal in N22, we found that $\hbar\Gamma_{23}\approx 200\,{\rm cm}^{-1}<\hbar(\Gamma_{g2}+\Gamma_{g3})$ (not shown). In Fig.~\ref{figureS6}b, the amplitude of the theoretical N22 is displayed as a function of the Huang-Rhys factor $S_2$, which is about $5\,\%$ of N11 over a range of realistic $S_2$ values. For a comparison, the Huang-Rhys factor $S_1$ of the vibrational mode with frequency $\hbar\nu_1\approx 668\,{\rm cm}^{-1}$ is marked by a vertical dashed line. These results imply that the small amplitude of the beating signal in N22 is mainly due to the high electronic decoherence rate of band 2.


\subsection{A correlated fluctuation model for bands 1 and 3 of C8O3}

Here we provide a correlated fluctuation model for bands 1 and 3 of C8O3 where coherent interaction between excitons and quasi-resonant vibrations is not considered. Within the level of Markovian quantum master equations, we show that the experimentally measured long-lived beating signals in N11 and R31 cannot be explained by correlated fluctuations.

The main idea of the correlated fluctuations is that when bands 1 and 3 are coupled to a common environment, the correlated noise enables the inter-exciton coherence $\ket{1}\bra{3}$ to decohere very slowly compared to the coherences $\ket{g}\bra{1}$ and $\ket{g}\bra{3}$ between electronic ground state and excitons. This is similar in spirit to the decoherence-free subspaces in quantum information theory\cite{Whaley_PRL1998_SI}. Here we consider a Markovian quantum master equation in the form of
\begin{equation}
	\frac{d}{dt}\rho(t)=-\frac{i}{\hbar}[\tilde{H}_e,\rho(t)]+\sum_{\omega}\sum_{\alpha,\beta}\gamma_{\alpha\beta}(\omega)\left(A_{\beta}(\omega)\rho(t)A_{\alpha}^{\dagger}(\omega)-\frac{1}{2}\{A_{\alpha}^{\dagger}(\omega)A_{\beta}(\omega),\rho(t)\}\right),
	\label{eq:RedfieldSecular}
\end{equation}
which is the same to Eq.~(3.143) in {\it The Theory of Open Quantum Systems} by H.-P. Breuer and F. Petruccione\cite{Breuer_SI}, which is called the Redfield equation with the secular approximation in some literature\cite{Rebentrost_JPCB2009_SI}. Here the interaction Hamiltonian is modeled by $H_{e-{\rm ph}}=\sum_{\alpha}A_{\alpha}\otimes B_{\alpha}$ with $A_{\alpha}=A_{\alpha}^{\dagger}$ and $B_{\alpha}=B_{\alpha}^{\dagger}$, each of which is a Hermitian operator of the system and environmental degrees of freedom, respectively. With the exciton states $\ket{k}$ defined by $\tilde{H}_{e}\ket{k}=\hbar\Omega_{k}\ket{k}$, we introduce a projection operator $\Pi(\Omega)=\sum_{\Omega=\Omega_k}\ket{k}\bra{k}=\sum_k \delta(\Omega,\Omega_k)\ket{k}\bra{k}$ where the Kronecker delta is defined by $\delta(i,j)=1$ if $i=j$ and $\delta(i,j)=0$ otherwise. In other words, $\Pi(\Omega)$ is a projection operator onto the exciton subspace belonging to the exciton energy $\Omega$. In Eq.~(\ref{eq:RedfieldSecular}), $A_{\alpha}(\omega)=\sum_{\Omega'-\Omega=\omega}\Pi(\Omega)A_{\alpha}\Pi(\Omega')=\sum_{\Omega,\Omega'}\delta(\omega,\Omega'-\Omega)\Pi(\Omega)A_{\alpha}\Pi(\Omega')$. The interaction Hamiltonian $H_{e-{\rm ph}}$ between excitons and background phonons is modeled by $A_{\alpha}=\ket{e_\alpha}\bra{e_\alpha}$ and $B_{\alpha}=\sum_{\xi}\hbar g_{\alpha\xi}(a_{\xi}^{\dagger}+a_{\xi})$, where $g_{\alpha\xi}$ denotes the coupling of the local excitation of site $\alpha$ to a background phonon mode $\xi$. When $g_{\alpha\xi}\neq 0$ and $g_{\beta\xi}\neq 0$ for different $\alpha$ and $\beta$, spatially separated sites $\alpha$ and $\beta$ are coupled to a common phonon mode $\xi$, leading to correlated fluctuations in the energy levels of the different sites $\alpha$ and $\beta$. The correlated fluctuations are absent when each site is coupled to an independent phonon bath, such that $g_{\alpha\xi}g_{\beta\xi}=0$ for all $\alpha\neq\beta$ and $\xi$. The information of the correlated fluctuations is included in the definition of $\gamma_{\alpha\beta}(\omega)$ in Eq.~(\ref{eq:RedfieldSecular})
\begin{equation}
	\gamma_{\alpha\beta}(\omega)=\frac{1}{\hbar^2}\int_{-\infty}^{\infty}ds e^{i\omega s}\mathinner{\langle B_{\alpha}^{\dagger}(s)B_{\beta}(0)\rangle},
\end{equation}
where for fixed $\omega$, $\gamma_{\alpha\beta}(\omega)$ form a positive matrix\cite{Breuer_SI}. Here $\gamma_{\alpha\beta}(\omega)=0$ for all $\alpha\neq\beta$ if each site is coupled to an independent phonon bath and $\gamma_{\alpha\beta}(\omega)\neq 0$ for some $\alpha\neq\beta$ if different sites $\alpha$ and $\beta$ are coupled to the same phonon modes.

Using experimentally measured absorption and 2D spectra of C8O3, we found that the electronic decoherence rate $\Gamma_{gk}$ of the coherence $\ket{g}\bra{k}$ between electronic ground state and band $k$ is given by $\hbar\Gamma_{g1}\approx 65\,{\rm cm}^{-1}$ and $\hbar\Gamma_{g3}\approx 150\,{\rm cm}^{-1}$ for bands 1 and 3, respectively. Within the level of the Markovian quantum master equation in Eq.~(\ref{eq:RedfieldSecular}), the decoherence rates $\Gamma_{g1}$ and $\Gamma_{g3}$ are given by
\begin{align}
	\Gamma_{g1}&=\frac{1}{2}\sum_{l\neq 1}\gamma_{1\rightarrow l}+\gamma_{g1},\label{eq:CDg1}\\
	\Gamma_{g3}&=\frac{1}{2}\sum_{l\neq 3}\gamma_{3\rightarrow l}+\gamma_{g3},\label{eq:CDg3}
\end{align}
where $\gamma_{k\rightarrow l}$ denotes the incoherent population transfer rate from band $k$ to band $l$, and $\gamma_{gk}$ represents the pure dephasing rate of the coherence $\ket{g}\bra{k}$. The population transfer rates $\gamma_{1\rightarrow l}$ and $\gamma_{3\rightarrow l}$ can be estimated using the exponential dynamics in 2D spectra. To estimate these rates, we performed a global target analysis on all parallel 2D spectra of C8O3~\cite{Milota_JPCA2013_SI}. We found that the population transfer rates from band 3 to lower energy bands 1 and 2 are approximately given by $\gamma_{3\rightarrow 1}\approx(300\,{\rm fs})^{-1}$ and $\gamma_{3\rightarrow 2}\approx(66\,{\rm fs})^{-1}$, corresponding to $\hbar\gamma_{3\rightarrow 1}\approx 18\,{\rm cm}^{-1}$ and $\hbar\gamma_{3\rightarrow 2}\approx 80\,{\rm cm}^{-1}$, respectively, and the other population transfer processes are slow in comparison, {\it i.e.}~$\gamma_{k\rightarrow l}\lesssim (2\,{\rm ps})^{-1}$. In this case, the pure dephasing rates of $\ket{g}\bra{1}$ and $\ket{g}\bra{3}$ are given by $\hbar\gamma_{g1}\approx \hbar\Gamma_{g1}\approx 65\,{\rm cm}^{-1}$ and $\hbar\gamma_{g3}\approx\hbar(\Gamma_{g3}-\frac{1}{2}\gamma_{3\rightarrow 1}-\frac{1}{2}\gamma_{3\rightarrow 2})\approx 101\,{\rm cm}^{-1}$, respectively.

The electronic decoherence rate $\Gamma_{13}$ of the inter-exciton coherence $\ket{1}\bra{3}$ between bands 1 and 3 is given by
\begin{equation}
	\Gamma_{13}=\frac{1}{2}\sum_{l\neq 1}\gamma_{1\rightarrow l}+\frac{1}{2}\sum_{l\neq 3}\gamma_{3\rightarrow l}+\gamma_{13},
	\label{eq:decay13}
\end{equation}
where $\gamma_{13}$ is the pure dephasing rate of the inter-exciton coherence in the presence of correlated fluctuations. We found that for given $\gamma_{g1}$ and $\gamma_{g3}$, the inter-exciton dephasing rate $\gamma_{13}$ should be higher than a theoretical lower bound given by
\begin{equation}
	\gamma_{13}\ge\left(\sqrt{\gamma_{g1}}-\sqrt{\gamma_{g3}}\,\,\right)^{2}.
	\label{eq:LowerBound}
\end{equation}
Within the level of the Markovian quantum master equation in Eq.~(\ref{eq:RedfieldSecular}), the lower bound is not violated by any spectral densities and correlated fluctuations, as will be shown below. Using the estimated values of the pure dephasing rates $\hbar\gamma_{g1}\approx 65\,{\rm cm}^{-1}$ and $\hbar\gamma_{g3}\approx 101\,{\rm cm}^{-1}$, we found that the lower bound in Eq.~(\ref{eq:LowerBound}) is reduced to $\hbar\gamma_{13}\gtrsim 4\,{\rm cm}^{-1}$. Therefore, even in the presence of correlated fluctuations, the decoherence rate $\Gamma_{13}$ of the inter-exciton coherence $\ket{1}\bra{3}$ in Eq.~(\ref{eq:decay13}) should be higher than a lower bound given by
\begin{equation}
	\Gamma_{13}\ge\frac{1}{2}(\gamma_{3\rightarrow 1}+\gamma_{3\rightarrow 2})+\left(\sqrt{\gamma_{g1}}-\sqrt{\gamma_{g3}}\,\,\right)^{2}\approx (100\,{\rm fs})^{-1}.
\end{equation}
It is notable that the lowest decoherence rate $\Gamma_{13}\approx (100\,{\rm fs})^{-1}$ ({\it cf}.~$\hbar\Gamma_{13}\approx 53\,{\rm cm}^{-1}$) is too high to explain the long-lived beating signals observed in the experiment, as shown in Fig.~\ref{figureS7}a, where the simulated results based on the correlated fluctuation model are shown as a blue solid line and the experimental results are shown as a light blue line. These results are mainly due to the fast population transfer from band 3 to bands 1 and 2 observed in the experiment.


\begin{figure}[tbp]
	\includegraphics{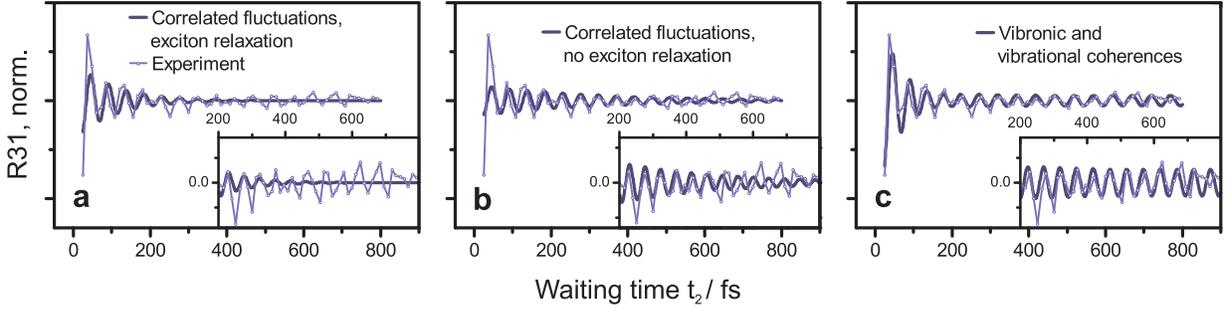}
	\caption{{\bf Correlated fluctuation model}. {\bf a}, Simulated beating signal for R31 in the presence of correlated fluctuations (without quasi-resonant vibrations). The modeled curve is shown as a solid blue line and experimental results are shown as a light blue line. Here we take the lowest decoherence rate $\Gamma_{13}\approx(100\,{\rm fs})^{-1}$ of the inter-exciton coherence allowed by correlated fluctuations, constrained by experimentally determined population transfer rates from band 3 to bands 1 and 2. As shown in the inset, correlated fluctuations cannot explain the experimentally measured long-lived beating signal in R31. {\bf b}, Simulated beating signal for R31 in the absence of the exciton relaxation, shown as a solid blue line. The correlated fluctuation model predicts the lowest decoherence rate $\Gamma_{13}\approx(303\,{\rm fs})^{-1}$ of the inter-exciton coherence when there is no exciton relaxation, even though this condition is not satisfied for C8O3. Nevertheless, the correlated fluctuation model cannot explain the experimentally measured long-lived beating signal in R31, which persist beyond $t_2\approx 800\,{\rm fs}$, as shown in an inset. {\bf c}, Simulated beating signal for R31 in the presence of a vibrational mode with frequency $\hbar\nu_1\approx 668\,{\rm cm}^{-1}$, which is quasi-resonant with the exciton energy splitting $\Delta\Omega_{31}$ between bands 1 and 3. Here the vibronic and vibrational coherences induce a long-lived beating signal in good agreement with the experimental results. The root-mean-square deviation (RMSD) of the experimental results and theoretical prediction in {\bf a}, {\bf b} and {\bf c} is 0.74, 0.86 and 0.59, respectively. We note that the correlated fluctuation model can also not explain the long-lasting beating signal in N11 (not shown).}
	\label{figureS7}
\end{figure}

We note that our results are not sensitive to the estimated values of the population transfer rates. The inter-exciton decoherence rate $\Gamma_{13}$ is minimized when there is no population transfer between excitons, {\it i.e.}~$\gamma_{k\rightarrow l}=0$ for all $k\neq l$, where the coherences between electronic ground state and excitons are destroyed only by pure dephasing noise, {\it i.e.}~$\hbar\Gamma_{g1}=\hbar\gamma_{g1}\approx 65\,{\rm cm}^{-1}$ and $\hbar\Gamma_{g3}=\hbar\gamma_{g3}\approx 150\,{\rm cm}^{-1}$. Even though this condition is not satisfied for C8O3, this is the best scenario of the correlated fluctuation model where the decoherence rate $\Gamma_{13}$ of the inter-exciton coherence is minimized
\begin{equation}
	\Gamma_{13}=\gamma_{13}\ge\left(\sqrt{\gamma_{g1}}-\sqrt{\gamma_{g3}}\,\,\right)^{2}\approx (303\,{\rm fs})^{-1}.
	\label{eq:optimal_correlated}
\end{equation}
However, even in this case, the lowest decoherence rate $\Gamma_{13}\approx (303\,{\rm fs})^{-1}$ supported by correlated fluctuations is not low enough to explain the experimentally measured long-lived beating signals in N11 and R31, which persist beyond $t_2\approx 800\,{\rm fs}$, as shown in Fig.~\ref{figureS7}b. This is due to the different decoherence rates $\Gamma_{g1}<\Gamma_{g3}$ of $\ket{g}\bra{1}$ and $\ket{g}\bra{3}$ observed in the experiment. This leads to a non-zero lower bound on the inter-exciton decoherence rate $\Gamma_{13}$, as shown in Eq.~(\ref{eq:optimal_correlated}). In addition, the beating signals in N11 and R31 consist of a short-lived component with 1/e decay time of $\sim$$\,66\,{\rm fs}$ as well as a long-lived component persisting up to $t_2\approx 1\,{\rm ps}$. This is contrary to the prediction of the correlated fluctuation model where a single oscillatory component is expected with 1/e decay time of $\Gamma_{13}^{-1}$. For a comparison, the theoretical prediction of the vibronic model is shown in Fig.~\ref{figureS7}c, where both short-lived and long-lived components are present. We note that in the vibronic model, the decay rate of the long-lived component is independent of $\Gamma_{g1}$ and $\Gamma_{g3}$, as it is determined by the other degrees of freedom, such as the dissipation rate $\gamma_{v}$ of the vibrations and the degree of vibronic mixing $\epsilon_2$ leading to a lifetime borrowing effect $\delta\gamma$, as shown in Eq.~(\ref{eq:exact_A(t2)}).

We now derive Eqs.~(\ref{eq:CDg1})-(\ref{eq:LowerBound}) using the Markovian quantum master equation in Eq.~(\ref{eq:RedfieldSecular}).

{\it (Dephasing noise)} We start with the case that $\omega=0$, leading to the pure dephasing noise. For the sake of simplicity, we assume that there is no degeneracy in the exciton energies $\Omega_k$, such that $\Omega_{k}\neq\Omega_{l}$ for all $k\neq l$. In this case, $A_{\alpha}(0)=\sum_{k}\ket{k}\bra{k}A_{\alpha}\ket{k}\bra{k}=\sum_{k}\abs{\left\langle k|e_{\alpha}\right\rangle}^{2}\ket{k}\bra{k}$. By substituting $\ket{g}\bra{1}$, $\ket{g}\bra{3}$ and $\ket{1}\bra{3}$ to the dissipator of the quantum master equation for $\omega=0$
\begin{equation}
	\frac{d}{dt}\rho(t)=\sum_{\alpha,\beta}\gamma_{\alpha\beta}(0)\left(A_{\beta}(0)\rho(t)A_{\alpha}^{\dagger}(0)-\frac{1}{2}\{A_{\alpha}^{\dagger}(0)A_{\beta}(0),\rho(t)\}\right),
	\label{eq:dissipator_d}
\end{equation}
one obtains the following pure dephasing rates of the coherences $\ket{g}\bra{1}$, $\ket{g}\bra{3}$ and $\ket{1}\bra{3}$
\begin{align}
	\gamma_{g1}
	&=\frac{1}{2}\sum_{\alpha,\beta}\abs{\left\langle 1|e_{\alpha}\right\rangle}^{2}\gamma_{\alpha\beta}(0)\mathinner{|\langle 1|e_{\beta}\rangle |}^{2}
	=\mathinner{|\vec{v}_{1}|}^{2},\\
	\gamma_{g3}
	&=\frac{1}{2}\sum_{\alpha,\beta}\abs{\left\langle 3|e_{\alpha}\right\rangle}^{2}\gamma_{\alpha\beta}(0)\mathinner{|\langle 3|e_{\beta}\rangle |}^{2}
	=\mathinner{|\vec{v}_{3}|}^{2},\\
	\gamma_{13}
	&=\frac{1}{2}\sum_{\alpha,\beta}\left(\abs{\left\langle 1|e_{\alpha}\right\rangle}^{2}-\abs{\left\langle 3|e_{\alpha}\right\rangle}^{2}\right)\gamma_{\alpha\beta}(0)\left(\mathinner{|\langle 1|e_{\beta}\rangle |}^{2}-\mathinner{|\langle 3|e_{\beta}\rangle |}^{2}\right)
	=\mathinner{|\vec{v}_{1}-\vec{v}_{3}|}^{2},
\end{align}
where $\vec{v}_k$ represents a vector defined by $\vec{v}_k=2^{-1/2}\hat{\gamma}^{1/2}\vec{w}_k$: here $\vec{w}_k$ is a real vector with elements $\abs{\left\langle k|e_{\alpha}\right\rangle}^{2}\ge 0$ and $\hat{\gamma}$ is a positive matrix\cite{Breuer_SI} with elements $\gamma_{\alpha\beta}(0)$, leading to to a positive matrix $\hat{\gamma}^{1/2}$ defined by $\hat{\gamma}=\hat{\gamma}^{1/2}\hat{\gamma}^{1/2}$. For given pure dephasing rates $\gamma_{g1}=\mathinner{|\vec{v}_{1}|}^{2}$ and $\gamma_{g3}=\mathinner{|\vec{v}_{3}|}^{2}$, the inter-exciton pure dephasing rate $\gamma_{13}$ is constrained by
\begin{equation}
	\gamma_{13}=\mathinner{|\vec{v}_{1}-\vec{v}_{3}|}^{2}\ge(\mathinner{|\vec{v}_{1}|}-\mathinner{|\vec{v}_{3}|})^{2}=\left(\sqrt{\gamma_{g1}}-\sqrt{\gamma_{g3}}\,\,\right)^{2},
	\label{eq:inequality}
\end{equation}
due to the triangle inequality, $\mathinner{|\vec{v}_{1}-\vec{v}_{3}|}+\mathinner{|\vec{v}_{3}|}\ge\mathinner{|\vec{v}_{1}|}$ and $\mathinner{|\vec{v}_{1}-\vec{v}_{3}|}+\mathinner{|\vec{v}_{1}|}\ge\mathinner{|\vec{v}_{3}|}$. Here the equality holds if and only if $\vec{v}_1$ is parallel to $\vec{v}_3$, which depends on $\hat{\gamma}^{1/2}$ (spectral densities and correlated fluctuations) as well as $\vec{w}_{1}$ and $\vec{w}_{3}$ (the spatial overlap between excitonic wavefunctions). Note that the lower bound in Eq.~(\ref{eq:inequality}) has been derived based on the positivity of $\hat{\gamma}^{1/2}$, which is satisfied for any spectral densities and correlated fluctuations. These results imply that the inter-exciton dephasing rate $\gamma_{13}$ can be reduced by the correlated fluctuations as well as the spatial overlap between excitonic wavefunctions.

{\it (Exciton relaxation)} We now consider the case that $\omega\neq 0$, leading to the incoherent population transfer between excitons. With $\Delta\Omega_{kl}=\Omega_k-\Omega_l$ denoting the exciton energy splitting between bands $k$ and $l$, $A_{\alpha}(\omega)=\sum_{k,l}\delta(\omega,\Delta\Omega_{kl})\ket{l}\bra{l}A_{\alpha}\ket{k}\bra{k}=\sum_{k,l}\delta(\omega,\Delta\Omega_{kl})\left\langle l|e_{\alpha}\right\rangle\left\langle e_{\alpha}|k\right\rangle\ket{l}\bra{k}$ and the dissipator of the quantum master equation for $\omega\neq 0$ is given by
\begin{align}
	\frac{d}{dt}\rho(t)&=\sum_{k\neq l}\sum_{\alpha,\beta}\gamma_{\alpha\beta}(\Delta\Omega_{kl})\left(A_{\beta}(\Delta\Omega_{kl})\rho(t)A_{\alpha}^{\dagger}(\Delta\Omega_{kl})-\frac{1}{2}\{A_{\alpha}^{\dagger}(\Delta\Omega_{kl})A_{\beta}(\Delta\Omega_{kl}),\rho(t)\}\right)\label{eq:dissipator_p},
\end{align}
where the population transfer rate $\gamma_{k\rightarrow l}$ from band $k$ to band $l$ is given by
\begin{equation}
	\gamma_{k\rightarrow l}=\sum_{\alpha,\beta}\left\langle k|e_{\alpha}\right\rangle\left\langle e_{\alpha}|l\right\rangle\gamma_{\alpha\beta}(\Delta\Omega_{kl})\mathinner{\langle l|e_{\beta}\rangle\langle e_{\beta}|k\rangle}\ge 0,
\end{equation}
which is positive, as $\gamma_{\alpha\beta}(\Delta\Omega_{kl})$ form a positive matrix\cite{Breuer_SI} for given $\Delta\Omega_{kl}$. By substituting $\ket{g}\bra{1}$, $\ket{g}\bra{3}$ and $\ket{1}\bra{3}$ to the dissipators in Eqs.~(\ref{eq:dissipator_d}) and (\ref{eq:dissipator_p}), one can show that the electronic decoherence rates $\Gamma_{g1}$, $\Gamma_{g3}$ and $\Gamma_{13}$ satisfy Eqs.~(\ref{eq:CDg1})-(\ref{eq:LowerBound}). These results are valid for any spectral densities and correlated fluctuations within the level of the Markovian quantum master equation in Eq.~(\ref{eq:RedfieldSecular}), which includes the theoretical models considered in previous studies\cite{Rebentrost_JPCB2009_SI}.

In summary, the correlated fluctuation model cannot explain the long-lived beating signals in N11 and R31 within the level of Markovian quantum master equations, as the decoherence rate $\Gamma_{13}$ of the inter-exciton coherence $\ket{1}\bra{3}$ is constrained by the experimentally observed asymmetric decoherence rates $\Gamma_{g1}$ and $\Gamma_{g3}$, {\it i.e.}~$\Gamma_{g3}\approx 2\Gamma_{g1}$, and the fast population transfer from band 3 to bands 1 and 2. We note that the asymmetric decoherence rates $\Gamma_{g1}$ and $\Gamma_{g3}$ are related to the fact that i) the lineshape of R31 is elongated along $\omega_1$-axis ({\it cf.}~Figures 2b and d in the main text), ii) the amplitudes of the short-lived beating signals in N11 and R31 are different in magnitude ({\it cf}.~Figures 3b and c in the main text), and iii) the absorptive linewidths of bands 1 and 3 are different ({\it cf.}~Fig.~\ref{figureS3} in the SI).

\end{widetext}


\end{document}